\documentclass[aps,showpacs,nofootinbib,floats,floatfix,preprintnumbers
, groupedaddress]{revtex4}
\pdfoutput=1
\usepackage{graphicx}
\usepackage{amsmath}
\usepackage{amsfonts}
\usepackage{amssymb}
\usepackage{xcolor, soul}
\usepackage{epstopdf}
\usepackage{float}
\usepackage{subfigure}
\usepackage{hyperref}
\usepackage{natbib}

\hypersetup{
  colorlinks   = true,
  citecolor    = red,
  linkcolor    = blue,
  urlcolor     = blue,
}
\def\beq{\begin{equation}}
\def\eeq{\end{equation}}
\def\bea{\begin{eqnarray}}
\def\eea{\end{eqnarray}}

\def\bse{\begin{subequations}}
\def\ese{\end{subequations}}

\def\wphi{w_{\rm re}}
\def\kf{k_{\rm end}}
\def\Tre{T_{\rm re}}

\def\Gev{\mbox{GeV}}
\def\mB{\mathcal{B}}
\def\mE{\mathcal{E}}
\def\pbi{\mathcal{P}_{\mathrm{B}}^{\mathrm{I}}}
\def\pei{\mathcal{P}_{\mathrm{E}}^{\mathrm{I}}}
\def\ei{\eta_{\mathrm{I}}}
\def\are{a_{\mathrm{re}}}

\def\rhoemi{\rho_{\mathrm{em}}^{\mathrm{I}}}
\def\Mp{M_{\mathrm{P}}}

\def\nw{n_{w}}
\def\nb{n_{\mathrm{B}}}
\def\ne{n_{\mathrm{E}}}
\def\Dneff{\Delta \mathrm{N}_{\mathrm{eff}}}
\def\nbe{n_{_{\mathrm{B}/\mathrm{E}}}}
\def\fnbe{\mathcal{F}_{n_{_{\mathrm{B}/\mathrm{E}}}}}

\def\HI{H_{_{\mathrm{I}}}}
\def\NI{N_{_{\mathrm{I}}}}
\def\kmin{k_{*}}
\def\Mpl{M_{_{\mathrm{P}}}}
\def\wre{w_{re}}
\def\Nre{N_{\mathrm{re}}}
\def\kre{k_{\mathrm{re}}}

\def\l{\left}
\def\r{\right}

\def\ei{\eta_{\mathrm{end}}}
\def\ere{\eta_{\mathrm{re}}}
\def\Pt{\mathcal{P}_{T}}

\def\Ptsi{\mathcal{P}_{T,s}^{\mathrm{inf}}}
\def\Ptvi{\mathcal{P}_{T,v}^{\mathrm{inf}}}

\def\Ptsre{\mathcal{P}_{T,s}^{\mathrm{re}}}

\def\nn{\nonumber}
\def\Gk{\mathcal{G}_k}
\def\Gki{\mathcal{G}_k^{\mathrm{inf}}}

\def\Fnb{\mathcal{F}_{n_B}(k)}
\def\pbi{\mathcal{P}_B^{\mathrm{inf}}}
\def\pei{\mathcal{P}_E^{\mathrm{inf}}}
\def\vk{\mathbf{k}}
\def\vq{\mathbf{q}}

\def\xre{x_{re}}
\def\xe{x_{\mathrm{end}}}
\def\Cmre{\mathcal{C}_{\mathrm{m,re}}}
\def\ogw{\Omega_{\mathrm{gw}}}
\def\ogwh{\Omega_{\mathrm{gw}}h^2}
\def\ogwp{\Omega_{\mathrm{gw}}^{\mathrm{pri}}}
\def\ogws{\Omega_{\mathrm{gw}}^{\mathrm{sec}}}

\def\ogwspi{\Omega_{\rm gw,pinf}^{\rm sec}}

\def\fne{\mathcal{F}_{n_{\rm E}}}

\def\hk{h_{\vk}}
\def\Ptp{\mathcal{P}^{\lambda}_{_{\rm PRI}}}
\def\Pts{\mathcal{P}^{\lambda}_{_{\rm SEC}}}
\def\Pt{\mathcal{P}^{\lambda}}
\def\ee{\eta_{\rm e}}
\def\Mp{M_{\rm P}}

\def\Ptpi{\mathcal{P}_{\rm inf}^{\rm PRI}}
\def\mIinf{\mathcal{I}_{\rm inf}}

\def\ere{\eta_{\rm re}}

\def\Ptrs{\mathcal{P}^{\lambda}_{_{\rm ra,SEC}}}

\def\hkre{h_{k,\rm re}^{\lambda}}
\def\mIra{\mathcal{I}_{\rm ra}}
\def\Fnb{\mathcal{F}_{n_{_{\rm B}}}}

\def\mCbe{\mathcal{C}_{_{\rm B/E}}}
\def\vk{\textbf{k}}
\def\vx{\textbf{x}}
\def\vq{\textbf{q}}
\def\Sem{\mathcal{S}_{\rm EM}}
\def\vx{\textbf{x}}
\def\gre{g_{\rm re}}
\def\ke{k_{\rm e}}
\def\f{\frac}
\def\Mpc{\mathrm{Mpc}}

\def\hkrep{h_{\vk,\rm re}^{\lambda '}}

\def\mS{\mathcal{S}}
\def\mH{\mathcal{H}}
\def\teta{\Tilde{\eta}}
\def\mGk{\mathcal{G}_{\vk}}
\def\pbe{\mathcal{P}_{\rm B/E}}
\def\fre{f_{\rm re}}
\def\mA{\mathcal{A}}
\def\mCe{\mathcal{C}_{\rm E}}
\def\mCb{\mathcal{C}_{\rm B}}
\def\ae{a_{\rm end}}

\def\mPt{\mathcal{P}_{\rm T}}

\graphicspath{{./figs/}}
\begin{document}

\title{Probing Reheating Phase via Non-Helical Magnetogenesis and Secondary Gravitational Waves\\
}
\author{Subhasis Maiti}
\email{E-mail: subhashish@iitg.ac.in}
\affiliation{Department of Physics, Indian Institute of Technology, Guwahati, 
Assam, India}
\author{Debaprasad Maity}
\email{E-mail: debu@iitg.ac.in}
\affiliation{Department of Physics, Indian Institute of Technology, Guwahati, 
Assam, India}

\author{Rohan Srikanth}
\email{E-mail: rohan.srikanth@uni-potsdam.de}
\affiliation{Institut f{\"u}r Physik und Astronomie, Universit{\"a}t Potsdam, Haus 28, Karl-Liebknecht-Str. 24/25, 14476, Potsdam, Germany}

\date{\today}

\begin{abstract}
In the past two decades, significant advancements have been made in observational techniques to enhance our understanding of the universe and its evolutionary processes. However, our knowledge of the post-inflation reheating phase remains limited due to its small-scale dynamics. Traditional observations, such as those of the Cosmic Microwave Background (CMB), primarily provide insights into large-scale dynamics, making it challenging to glean information about the reheating era.
In this paper, our primary aim is to explore how the generation of Gravitational Waves (GWs) spectra, resulting from electromagnetic fields generated during inflation, can offer valuable insights both into the magnetogenesis model and Reheating dynamics. We investigate how the spectral shape of GWs varies across different frequency ranges, depending on the model parameters, the magnetogenesis parameters, and reheating dynamics. For this, we consider a well-known non-helical magnetogenesis model, where the usual electromagnetic kinetic term is coupled with a background scalar. Notably, for such a scenario, we observe distinct spectral shapes with sufficiently high amplitudes for different reheating histories with the equation of state parametrized by ($w_{\rm re}$). We identify spectral breaks in the GW spectra for both $w_{\rm re}<1/3$ and $w_{\rm re}>1/3$ scenarios.
We find that future GW experiments such as BBO, LISA, SKA, and DECIGO are well within the reach of observing those distinct spectral shapes and can potentially shed light on the underlying mechanism of the reheating phase. 
Finally, we attempt to constrain the non-helical models under consideration utilizing the latest NANOGrav 15-year observation campaign. We find that in order to explain the observed GW spectrum in the nano-Hz range, the required magnetic field spectrum needs to be very stiff. However, such a stiff magnetic field spectrum in turn yields a strongly blue-tilted GW spectrum, which conflicts with other cosmological observations.
\end{abstract}
\maketitle

\section{Introduction:}
The last decade can be called \textit{the} era of Gravitational waves (GW) in the field of observational cosmology. Despite its long-standing prediction undergoing painstaking conceptual checks in Einstein's General Relativity framework, detection of GW was close to impossible till the LIGO-Virgo collaboration detected the GW signal from a distant source of merging black hole binary in 2015 \cite{LIGOScientific:2016aoc, LIGOScientific:2016vlm, LIGOScientific:2016emj, LIGOScientific:2016vbw, LIGOScientific:2017bnn}. Subsequently, within ten years, a large number of GW sources have been detected along with the latest revelation of stochastic GW background in the Nano-Hertz frequency range decoded from the measured tiny variation of the periods of Pulsar Timing Arrays (PTA) scattered in the sky \cite{NANOGrav:2023gor, Antoniadis:2023utw, Reardon:2023gzh, Zic:2023gta, Xu:2023wog}. Except for different classes of precisely known astrophysical binary sources, stochastic Nano-Hz GW sources are not easy to identify. And this certainly signals the beginning of an era of GWs starting to play their long-expected role in revealing the mystery of the early universe. 

The mechanism of producing stochastic GW has been the subject of intensive theoretical investigation over several decades in theoretical cosmology. Primordial stochastic gravitational waves (PSGWs) are an ideal probe for studying the early universe's dynamics and microscopic physics due to their pristine nature. These waves are hypothesized to have originated from the quantum vacuum and amplified by the inflationary mechanism. The detection of PSGWs would not only validate the inflationary mechanism but also provide crucial insights into physics at extremely high energy scales. PSGWs are generated from the inflationary quantum vacuum, and their production and evolution have been extensively studied in the literature \cite{Starobinsky:1979ty, Grishchuk:1974ny, Guzzetti:2016mkm, Haque:2021dha}.

Inflation is a well-established mechanism in which spacetime vacuum fluctuations are amplified, leading to the generation of tensor perturbations without the need for external sources. The gravitational waves (GWs) produced from these vacuum fluctuations are referred to as primary gravitational waves (PGWs) \cite{Starobinsky:1979ty, Grishchuk:1974ny, Guzzetti:2016mkm, Haque:2021dha}. In our model, electromagnetic (EM) fields are also generated during inflation. It is well known that such EM fields can act as efficient sources of secondary gravitational waves (SGWs) \cite{ PhysRevD.65.023517, Sorbo:2011rz, Caprini:2014mja, Ito:2016fqp, Sharma:2019jtb, Okano:2020uyr, PhysRevD.64.123514, Di:2017ndc, Fu:2019vqc, PhysRevD.103.083510, Bhaumik:2020dor, Solbi:2021wbo, Figueroa:2021zah, Maiti:2024nhv}.
Unlike PGWs, SGWs carry imprints not only of the background dynamics but also of the intrinsic properties of their sources. In this work, we focus on the production and evolution of SGWs sourced by inflationary electromagnetic fields, with particular emphasis on the effects of the reheating phase. Our analysis considers a specific class of electromagnetic sources assumed to originate from inflationary magnetogenesis.

Studies on inflationary magnetogenesis \cite{Ratra:1991bn, Demozzi:2009fu, Kandus:2010nw, PhysRevD.37.2743, Durrer:2013pga, Ferreira:2013sqa, Subramanian:2015lua, Kobayashi:2014sga, Haque:2020bip, Tripathy:2021sfb, Li:2022yqb, Adshead:2016iae, Campanelli:2008kh, Jain:2012jy, Caprini:2014mja, Bamba:2021wyx, Maiti:2025rkn, Papanikolaou:2024cwr} has mostly focused on producing large-scale magnetic field (LSMF) in the intergalactic medium in voids ($\sim 10^{-16}$ Gauss of coherence length as large as Mpc scales) as hinted by different observations \cite{neronov2010evidence, Essey:2010nd, PhysRevD.98.083518}. Furthermore, such large-scale magnetic fields are also argued to act as a tiny seed initial magnetic field which may be amplified by the well-known galactic dynamo mechanism  \cite{2010ASPC..438..197F,beck2016magnetic, Haverkorn:2008tb,kronberg2001magnetic}
to produce magnetic fields at various astrophysical and galactic scales of order a few micro Gauss \cite{Grasso:2000wj, kronberg2001magnetic, PhysRevD.37.2743}. Non-linear magneto-hydrodynamic effects at astrophysical and galactic scales generically erase the detailed properties of the initial seed field, which makes it difficult to verify such inflationary mechanisms of magnetogenesis models. Moreover, magnetic fields in the voids of the Mpc scale are extremely challenging to detect. Apart from the indirect $\gamma$-ray observational lower limit of LSMF setting some constraints \cite{MAGIC:2022piy, takahashi2011lower, Arlen_2014}, all the magnetogenesis models are largely unconstrained and poorly understood from the observational point of view.
Therefore, it is essential to look for some indirect observables that can help understand such mechanisms better. In this realm, studies have been performed on the physical effect of LSMF on various important cosmological observables such as inflationary power spectrum \cite{Hortua:2014wna}, CMB \cite{Paoletti:2022gsn, Zucca:2016iur}, gravitational waves \cite{Sorbo:2011rz, PhysRevD.65.023517, Ito:2016fqp, Sharma:2019jtb, Okano:2020uyr}. In this paper, we shall study in detail the production of SGW due to the electromagnetic field produced during inflation. As we shall observe, the SGW dynamics and its spectrum crucially depend on the magnetogenesis model and the background evolution. Therefore, SGW naturally encodes the physics of inflation and the magnetogenesis models together.

After the end of inflation \cite{Guth:1980zm, Linde:1981mu, Albrecht:1982wi, Starobinsky:1980te, PhysRevD.50.7222}, both GWs and the produced electromagnetic field will evolve through a reheating phase which depends on the very nature of the inflaton field near its minimum and how it is coupled to the standard model fields. In the simplest perturbative framework, this phase is parametrized by the reheating temperature $(\Tre)$ which encodes inflaton decay width, $\Gamma$ into standard model fields through the relation $\Tre \propto \sqrt{\Gamma}$, and the inflation equation of state ($\wre$) which encodes the nature of the inflaton potential \cite{Kolb:1990vq, Kofman:1997yn, RevModPhys.78.537, Allahverdi:2010xz, Amin:2014eta, Drewes:2017fmn, Haque:2022kez, Adhikari:2019uaw, Yogesh:2024mpa, Ghosh:2024ybs, Gangopadhyay:2022vgh,Khan:2022odn}.
Even though the reheating temperature is broadly bounded within $10^{15} \mbox{GeV} > \Tre > T_{BBN} \sim 10~ \mbox{MeV}$ \cite{Kawasaki:1999na, Cook:2015vqa, PhysRevD.107.043531, Hasegawa:2019jsa}, the inflaton/reheating equation of state is mostly unconstrained (see recent works on this \cite{PhysRevD.107.043531}). A recent surge of activity on understanding this phase has been to realize the underlying connection among inflationary parameters such as scalar spectral index $(n_s)$ and the reheating parameters $(\Tre, \wre)$\cite{PhysRevLett.113.041302, PhysRevD.82.023511, PhysRevD.98.103525, PhysRevD.107.043531, Adhikari:2019uaw, Yogesh:2024mpa, Yogesh:2025wak}.
Furthermore, the evolution of PGW has also been shown to encode valuable information when passing through this phase, see \cite{Haque:2021dha} and references therein.
In this paper, we will analyze in detail the production of SGW, a specific class of non-helical magnetogenesis scenario accounting for the observable constraints on the present-day LSMF combined with the CMB anisotropies, BBN, and other sensitivity curves of the future observatories. Our SGW analysis is observed to put interesting constraints on the nature of inflation through its equation of state $\wre$, reheating temperature $\Tre$, and the inflationary magnetogenesis model itself. To generate the LSMF, we consider a simple model of primordial magnetogenesis with the coupling $f^2(\phi)FF$ \cite{Subramanian:2015lua, Kobayashi:2014sga, Haque:2020bip, Tripathy:2021sfb, Dimopoulos:2024jnv}.

The structure of the paper is organized as follows. In Section~\ref{sec2}, we present a detailed analysis of the primordial magnetic fields generated during inflation, emphasizing the impact of non-trivial reheating dynamics. In particular, we investigate how the presence of conductivity during reheating can significantly influence the present-day strength of magnetic fields on large scales. In Section~\ref{sec3}, we explore how the presence of electromagnetic fields modifies the gravitational wave (GW) spectrum under generalized reheating dynamics, and discuss the prospects for detecting these signals with future GW observatories. Section~\ref{sec4} is devoted to presenting key results from our analysis of the GW spectral features, along with constraints on the reheating dynamics derived from both the tensor-to-scalar ratio bounds and the limits on $\Dneff$ bound. Finally, in Section~\ref{sec5}, we summarize our main findings and conclude.

\section{Magnetogenesis: quantizing the gauge field and evolution of its power spectrum}\label{sec2}
To make our paper self-contained, we briefly describe the quantization of gauge fields and define the associated spectrum in the Friedmann–Lemaître–Robertson–Walker (FLRW) background. Maxwell's theory in four dimensions is conformally invariant, which means that to generate magnetic fields in a spatially flat universe as discussed in \cite{Kobayashi:2014sga, Ferreira:2013sqa, Subramanian:2015lua, Haque:2020bip, Tripathy:2021sfb, Campanelli:2008kh, Jain:2012jy, Caprini:2014mja, Bamba:2021wyx}, one must break conformal invariance. To achieve this, a time-dependent effective gauge coupling $f(\eta)$ is typically introduced \cite{Kobayashi:2014sga,Ferreira:2013sqa,Subramanian:2015lua,Haque:2020bip,Tripathy:2021sfb}.
\begin{align}
\Sem=
-\frac{1}{4}\int d^4x\sqrt{-g}g^{\alpha\beta}g^{\mu\nu}f^2(\eta)F_{\mu\alpha}F_{\nu\beta} .\label{eq:action_1}
\end{align}
Where $F_{\mu\nu}=\partial_{\mu}A_{\nu}-\partial_{\nu}A_{\mu}$, is the electromagnetic field tensor and $A_{\mu}$ is four-vector potential. The possible source of the gauge coupling $f(\eta)$ will not be discussed here and will be taken up in our future publication. Here, we will follow the quantization procedure discussed in the paper \cite{Maity:2021qps}.
Throughout this analysis, we will treat electromagnetic field perturbatively compared to inflation and radiation energy density, i.e., ${\rho_{em}}/{\rho_{inf}}, {\rho_{em}}/{\rho_{rad}}<1$, and spatially Flat FLRW metric background in conformal coordinates is
\begin{align}\label{metric-element}
ds^2=g_{\mu\nu}dx^{\mu}dx^{\nu}=a^2(\eta)(-d\eta^2+dx^2),
\end{align}
where `$\eta$' is the conformal time. Spatial flatness helps us to denote the vector potential in terms of irreducible components as
$
A_\mu=\left(A_0,\partial_iS+ A_i\right)$ with the traceless condition $ \delta^{ij}\partial_i A_j=0.$ In terms of these components the action Eq.\eqref{eq:action_1} becomes,
\begin{equation}
\label{action_pot}
\Sem=\frac{1}{2}\int d\eta d^3x f(\eta)^2 (A_i^\prime {A^i}^\prime-\partial_i A_j\partial^i A^j).
\end{equation}
Because of the inherent conformal invariance, the action becomes independent of the scale factor. The spatial index will be raised or lowered by the usual Kronecker delta function. Assuming the Fourier expansion of $A_i$ as
\begin{equation}
\label{mode_EM}
A_i(\eta,\vx)=\sum_{\lambda=1,2}\int\frac{d^3k}{(2\pi)^3}\epsilon^{\lambda}_i(\textbf{k})  e^{i\textbf{k.x}}u_k^{\lambda}(\tau),
\end{equation}
with the reality condition $u^{\lambda}_{-\vk} = u^{\lambda*}_{\vk}$, the polarization vector $\epsilon_i^{\lambda}(\vk)$ corresponding to two modes $\lambda=1,2$, and satisfies
$
\epsilon_i^{\lambda}(\vk) k_i=0$ and $\epsilon_i^{\lambda}(\vk)\epsilon_i^{\lambda'}(\vk)=\delta_{\lambda\lambda'}.
$
The associated mode function satisfies \cite{Haque:2020bip,Tripathy:2021sfb},
\beq\label{mode_eq1}
u^{\lambda\prime\prime}_{\vk}+ 2\frac{f^\prime}{f}u^{\lambda \prime}_{\vk}+k^2u^{\lambda}_{\vk}=0 ,
\eeq
where the prime denotes the derivative with respect to the conformal time $\eta$. 
The electric and magnetic power spectrum in terms of the mode function is written as, 
\begin{align}\label{eq9}
\mathcal{P}_E=\frac{f^2 k^3}{4\pi^2a^4}\sum_{\lambda=\pm} |u_\vk^{\lambda'}|^2~~;~~
\mathcal{P}_B=\frac{f^2 k^5}{4\pi^2a^4}\sum_{\lambda=\pm} |u_\vk^{\lambda}|^2 .
\end{align}
Power spectrum can now be computed once we know the mode function $u_\vk^{\lambda}$ for a given inflationary and magnetogenesis model. Since, $(\mathcal{P}_E,\mathcal{P}_B)$ are connected to the direct physically measurable quantities, we chose to take a slightly different but transparent approach by directly solving the equations for comoving power spectrum $(\mathcal{P}^c_E = ({a^4}/{f^2}) \mathcal{P}_E,\mathcal{P}^c_B = ({a^4}/{f^2}) \mathcal{P}_B)$ as
\bea \label{eq10}
\mathcal{P}^{c'}_E + 4 \frac {f'}{f} \mathcal{P}^c_E = -\mathcal{P}^{c'}_B \\\nonumber
\mathcal{P}^{c''}_B + 2 \frac {f'}{f} \mathcal{P}^{c'}_B + 2 k^2 \mathcal{P}^c_B   = 2 k^2 \mathcal{P}^c_E.
\eea
The above equations clearly state that once the comoving magnetic power spectrum $\mathcal{P}^c_B$ is known, $\mathcal{P}^c_E$ can be automatically obtained from the second equation, and vice versa. Hence, it is sufficient to set the initial condition for the comoving magnetic spectrum, $\mathcal{P}^c_B|_{-k\tau\rightarrow \infty} = {k^4}/{4 \pi^2 f^2}$, which can be obtained by explicitly solving for the electromagnetic mode function, (Eq.\ref{mode_eq1}) for $u_\vk^{\lambda}$, and applying both the Bunch-Davis (BD) vacuum and normalization conditions, namely $\l. u_\vk^{\lambda}\r|_{\rm BD}={1}/({\sqrt{2k f^2}})e^{-ik\eta}$.

\subsection{Defining Reheating Temperature and Inflationary Parameters} \label{subsec:inflation-model}

In the standard cosmological model, inflation is followed by a phase of reheating, during which energy is transferred from the inflaton field to radiation. This phase is characterized by the mass of the inflaton, its self-coupling, and its coupling to radiation. In the perturbative regime, these parameters can be related to the equation-of-state (EoS) parameter~$\wre$ and the reheating temperature~$\Tre$.

During reheating, the energy density of the inflaton evolves as $a^{-3(1+\wre)}$ since the inflaton field dominates this phase. The reheating temperature, $\Tre$, is defined as the point at which the energy densities of the inflaton ($\rho_\phi$) and radiation ($\rho_{\rm r}$) become equal. To maintain a model-independent approach, we consider the EoS $(\wre)$ and the reheating temperature $(\Tre)$ as free parameters. 

The duration of the reheating period can be quantified by the total number of e-folds during reheating, denoted as $N_{\rm re}$, and is given by \cite{PhysRevLett.113.041302, Chakraborty:2024rgl}
\begin{align}
    \Nre = \frac{1}{3(1+\wre)}\ln\left(\frac{90\HI^2\Mp^2}{\pi^2\gre\Tre^4}\right).\label{eq:nre}
\end{align}
Here, $\HI$ 
is the Hubble parameter during inflation, assumed to remain constant throughout the inflationary era. The reduced Planck mass is $\Mp = {1}/{\sqrt{8\pi G}} \simeq 2.14 \times 10^{18}$ GeV, and $\gre \simeq 106.7$ represents the number of relativistic degrees of freedom at the beginning of the radiation-dominated era. 

We can also define other relevant parameters as follows \cite{Chakraborty:2024rgl}:
\begin{align}
    \kre \simeq 3.9 \times 10^6 \left(\frac{\Tre}{10^{-2}\, \text{GeV}}\right) \, \mathrm{Mpc}^{-1},
\end{align}
where $\kre$ represents the lowest mode that can re-enter the horizon before the end of reheating.

Additionally, we define $\ke$ as \cite{Chakraborty:2024rgl}
\begin{align}
    \ke = \left(\frac{43 \gre}{11}\right)^{1/3} \left(\frac{\pi^2 \gre}{90}\right)^{\alpha} \frac{\HI^{1 - 2 \alpha} \Tre^{4 \alpha - 1} T_0}{\Mp^{2 \alpha}},\label{eq:def_ke}
\end{align}
Here $\ke$ is the highest mode that can leave the horizon at the end of inflation. In the above we have defined $\alpha ={1}/{3(1 + \wre)}$ and $T_0 = 2.725 \, \text{K}$ is the present-day CMB temperature. Here, we consider $a_0 = 1$ as the present-day value of the scale factor.

\subsection{Evolution of electromagnetic power spectrum through different phases}
Our main objective is to trace the evolution of the electromagnetic power spectrum from the inflationary period to the current era and then compare it with the observed limit on the magnetic power spectrum. To this end, we will examine the gravitational wave power spectrum's development in the presence of this electromagnetic source.

\subsubsection{Evolution During inflation: } During Inflation, we assume the ansatz for the gauge coupling function
as \cite{Kobayashi:2014sga,Haque:2020bip,Tripathy:2021sfb}
\begin{align}\label{eq11}
  f(\eta)=\left\{ \begin{matrix}
  \left(\frac{a}{\ae}\right)^n &  \mbox{    for $a\leq \ae$}\\
  1  & \mbox{    for $a>\ae$}
  \end{matrix}\right.  .
\end{align}
Note that after the end of inflation set at $\ae$, the coupling function becomes unity, which restores the usual conformal symmetry of the electromagnetic field.

To solve for the spectrum analytically, we further assumed inflation to be approximately de Sitter type, and with this assumption, the scale factor behaves as $a=-1/\HI\eta$, where $\HI$ is the Hubble constant. With the above choice of gauge coupling function, one can easily obtain the analytic solution for the magnetic power spectrum as
\bea
\mathcal{P}^c_B = ({a^4}/{f^2}) \mathcal{P}_B = (-k\eta)^{1 + 2n}(c_1 H_{n +\frac 12}(k\eta)^2 + c_2 |H_{n +\frac 12}(k\eta)|^2 + c_3 H^*_{n+\frac 12}(k\eta)^2 ).
\eea
Where ${c_1,c_2,c_3}$ are the integration constants, and $(H_{n +1/2}, H^*_{n + 1/2})$ are the Hankel functions of the first kind and their complex conjugate. The electric power spectrum can be easily computed from Eq.(\ref{eq10}). Since, in the asymptotic past limit, $a^4\mathcal{P}^c_B|_{-k\eta\rightarrow \infty} = { k^4}/{4 \pi^2 f^2}$, one obtains the integration constants to be
\bea\label{13}
c_1=c_3 =0 ~;~ c_2 = \frac {k^4} {8 \pi} \left(\frac{1}{-k\ee}\right)^{2n}.
\eea
Once we obtain the expression for $\mathcal{P}^c_B$, the corresponding electric counterpart can be obtained from Eq.(\ref{eq10}), and using Eq. (\ref{13}), we get the final expressions for the electromagnetic power spectrum \cite{Haque:2020bip} as, 
\begin{align}\label{eq14}
 \pbi(k,\eta)=\frac{d\rho_B}{dln(k)}=\frac{k^4}{8\pi a^4(\eta)}(-k\eta)\left|H^{(1)}_{n+\frac{1}{2}}(-k\eta)\right|^2 ,
\end{align}
\begin{align}\label{eq15}
\pei(k,\eta)=\frac{d\rho_E}{dln(k)}=\frac{k^4}{8\pi a^4(\eta)}(-k\eta)\left|H^{(1)}_{n-\frac{1}{2}}(-k\eta)\right|^2  .
\end{align}
To analyse the power spectrum's characteristics at the super-horizon scale, where all the relevant modes are well outside the Hubble radius at the end of inflation, we consider the limit where ${k}/{a\HI}<<1$. Depending on the sign of $n$, we can express the electric and magnetic power spectrum in a simpler form, as explained in \cite{Subramanian:2015lua, Haque:2020bip, Tripathy:2021sfb}
\begin{align}\label{eq:pbi_1}
\left.  \begin{matrix}
\pbi(k,\eta)=\frac{\HI^4}{8\pi}\frac{2^{2|n|+1}\Gamma^2\left(|n|+\frac{1}{2}\right)}{\pi^2}(-k\eta)^{-2|n|+4}\\
\pei(k,\eta)=\frac{\HI^4}{8\pi}\frac{\Gamma^2\left(|n|-\frac{1}{2}\right)2^{2|n|-1}}{\pi^2}(-k\eta)^{-2|n|+6}
  \end{matrix}\right\} \mbox{     for $n>\frac{1}{2}$},
\end{align}
\begin{align}\label{eq:pbi_2}
\left.  \begin{matrix}
\pbi(k,\eta)=\frac{\HI^4}{8\pi}\frac{\Gamma^2\left(|n|-\frac{1}{2}\right)2^{2|n|-1}}{\pi^2}(-k\eta)^{-2|n|+6}\\
\pei(k,\eta)=\frac{\HI^4}{8\pi}\frac{2^{2|n|+1}\Gamma^2\left(|n|+\frac{1}{2}\right)}{\pi^2}(-k\eta)^{-2|n|+4}
  \end{matrix}\right\} \mbox{     for $n<-\frac{1}{2}$} .
\end{align}
After obtaining the power spectrum during inflation, we can use it as a boundary condition to solve for the power spectrum during reheating. It is important to note that for $n > 1/2$, the ratio of the two field strengths turns out to be ${\pbi(k)}/{\pei(k)} \propto (k\eta)^{-2} \gg 0$.
This immediately suggests that, at the end of inflation, the magnetic field strength dominates over the electric field strength for $n > 1/2$. However, it should be emphasized that the case $n > 0$ suffers from a strong coupling problem. On the other hand, for $n < 0$, the situation is reversed: the electric field strength dominates over the magnetic field strength, i.e. ${\pei(k)}/{\pbi(k)} \propto (k\eta)^{-2} \gg 0$ and in this case there is no strong coupling issue.
To estimate the magnetic field strength at the end of inflation for a specific set of parameters, we consider an inflationary energy scale of $\HI\simeq 10^{-5} \Mp$ and an inflationary duration corresponding to $\NI=55$ e-folds.Now we can define the magnetic field strength at the end of inflation as $B(k,\ee)=\sqrt{\pbi(k,\ee)}$. For $n=2$, which corresponds to a scale-invariant magnetic spectrum, the field strength at a comoving scale of $ k = 1~\Mpc^{-1}$ is found to be $B_{n=2}(1~\text{Mpc}^{-1}) \simeq 10^{46}$~\text{G}.
In contrast, for $n=-2$, where the magnetic spectrum exhibits a scale-dependent behavior with $B_{n=-2} \propto k$, the field strength at the same scale is estimated as $B_{n=-2}(1~\Mpc^{-1}) \simeq 10^{23}$~\text{G}.  
Notably, for scale-dependent spectra, the field strength is influenced by the duration of inflation due to the factor $ (k/\ke)^{n_B} $, whereas for scale-invariant spectra, it remains independent of the inflationary duration. 
\paragraph{\underline{Results for Inflationary Magnetogenesis:}}
The strength of the electromagnetic power spectrum at the end of inflation is governed by equations (\ref{eq:pbi_1}),(\ref{eq:pbi_2}). Assuming an instantaneous reheating scenario, and the inflationary Hubble parameter $\HI \simeq 10^{-5} \Mp$, we can determine the present-day magnetic field strength as follows:
\begin{align}
    B_0(k)\simeq 1.84 \times 10^{-12}\mbox{ G}\l(\frac{k}{\kf}\r)^{2-|n|}\times\l\{\begin{matrix}
       2^{n+\frac12}\Gamma(n+ \frac12)  
       & \mbox{ for $n>1/2$}\\
        2^{n-\frac12}\Gamma(n-\frac12)
        \frac{k}{\kf} & \mbox{ for $n<-1/2$}
    \end{matrix}\r. .
\end{align}
Notably, the case of $n=2$ yields a scale-invariant magnetic field. Despite the scale-invariance, it faces the challenge of the strong coupling problem due to the large effective electromagnetic gauge coupling $\propto 1/f^2$, at the onset of inflation. However, it can generate a magnetic field strength of around $\sim 10^{-11}$ G. Note that such a strong magnetic field can have a sizable impact on the tensor-to-scalar ratio, which we will discuss in the GW section. On the other hand, the $n=-2$ coupling leads to a scale-dependent magnetic field strength is $B_0(1\,\Mpc^{-1})\simeq1.45\times 10^{-36}$ G, which is very small compared with the current observational bound on the 1 Mpc scale \cite{neronov2010evidence, Essey:2010nd, PhysRevD.98.083518}.

\subsubsection{Backreaction and Strong Coupling Problem}
In models featuring a coupling of the form $ f^2 F_{\mu\nu}F^{\mu\nu} $, the effective electromagnetic gauge coupling is given by $ \alpha_{\rm em} = e^2 / f^2 $. Consequently, if the coupling function satisfies $ f \ll 1 $ at the beginning of inflation, the effective gauge coupling becomes very large. This places the system in the strong coupling regime, rendering the perturbative treatment of the gauge field unreliable. Conversely, if the scenario is initialized with $ f \gg 1 $ and evolves toward $ f = 1 $ by the end of inflation, thus restoring conformal symmetry, the effective gauge coupling remains small throughout, with $ \alpha_{\rm em} \lesssim e^2 $. In such cases, the model avoids the strong coupling problem.

For the specific choice of the coupling function discussed in Eq.~\eqref{eq11}, we find that for $ n > 0 $, the initial value of the function is very small, $ f(\eta_i) \ll 1 $, where $ \eta_i $ denotes the beginning of the inflationary era. This implies that models with $ n > 0 $ generally suffer from strong coupling issues at early times. In contrast, for $ n < 0 $, the coupling function starts with a large value, avoiding the strong coupling problem altogether. As discussed in Eq.~\eqref{eq:pbi_2}, for the regime $ n < -1/2 $, the electric and magnetic power spectra exhibit the following behavior
\begin{align}
\pbi(k,\eta)=\frac{\HI^4}{8\pi}\frac{\Gamma^2\left(|n|-\frac{1}{2}\right)2^{2|n|-1}}{\pi^2}(-k\eta)^{\ne+2},\label{eq:pbi_adibatic}\\
\pei(k,\eta)=\frac{\HI^4}{8\pi}\frac{2^{2|n|+1}\Gamma^2\left(|n|+\frac{1}{2}\right)}{\pi^2}(-k\eta)^{\ne},
\end{align}
where $ \ne = 4 - 2|n| $ denotes the spectral index of the electric field. From the above discussion, it is evident that a scale-invariant magnetic power spectrum requires $ n = -3 $. For this specific choice of the coupling parameter, the electric spectrum scales as $ \mathcal{P}_E(k,\eta) \propto (-k\eta)^{-2} $, implying that in the limit $ |-k\eta| \to 0 $, the energy density of the electric field diverges. Consequently, the electric energy density may exceed the inflationary background energy density by the end of inflation. Therefore, the choice $ n = -3 $ leads to a significant backreaction problem.

This consideration imposes both upper and lower bounds on the coupling parameter $ n $. Beyond a certain threshold, the backreaction becomes negligible and the model remains consistent. To determine the lower bound on $ n $, one must compute the total energy density associated with the electromagnetic field at the end of inflation, i.e., at conformal time $ \eta = \ee $. For $ n < 0 $, the total energy density of the electromagnetic field is given by
\begin{align}
    \rhoemi=\int_{k_*}^{\ke}d\ln(k) \times \mathcal{P}_{\rm B/E}(k,\ee) \simeq \frac{\HI^4}{8\pi} \frac{2^{2|n|+1}\Gamma^2\l(|n|+\frac{1}{2}\r)}{\pi^2} \frac{\l(1-(k_*/\ke)^{\ne}\r)}{\ne} .
\end{align}
To compute the total electromagnetic energy density, we integrate over the comoving wavenumber $ k $ in the range $ k_{*} \leq k \leq k_{\rm end} $, where $ k_{*} $ is the pivot scale corresponding to large wavelengths observable in the CMB, and $ k_{\rm end} $ is the smallest scale that exits the horizon at the end of inflation. We take the ultraviolet cutoff to be $ k_{\rm UV,c} = k_{\rm end} $, as only those modes that exit the horizon during inflation are excited via the coupling; modes that remain subhorizon throughout inflation stay in the Bunch-Davies vacuum and do not contribute to the observable magnetic field strength.

By requiring that the total energy density of the electromagnetic field remains subdominant compared to the background inflationary energy density at the end of inflation, we find that the coupling parameter must satisfy $ n \geq -2.2 $, assuming an inflationary energy scale $ H_{\rm inf} \simeq 10^{-5} \Mp $ and an inflationary duration of $ N_{\rm inf} \simeq 55 $, where $ N_{\rm inf} $ denotes the number of e-folds between horizon exit of the pivot scale and the end of inflation. Thus, to simultaneously avoid both the backreaction and strong coupling problems, the allowed range for the coupling parameter is $ -2.2 \leq n < 0 $.
Within this range, the magnetic power spectrum is always blue-tilted. For instance, at $ n = -2 $, the electric power spectrum becomes scale-invariant, while the magnetic power spectrum scales as $ \mathcal{P}_B(k, \ee) \propto k^2 $, indicating that most of the magnetic energy is concentrated near the cutoff scale $ k \simeq k_{\rm end} $. As a result, the amplitude of large-scale magnetic fields is suppressed at the end of inflation. 


In the following section, we explore how reheating scenarios involving negligible or low electrical conductivity (EC) can enhance the magnetic field through induction effects (see, for instance,~\cite{Kobayashi:2019uqs}).

\subsection{Evolution During reheating phase:}
During reheating, we now consider two extreme cases with regard to the electrical conductivity of the background medium. 
\begin{enumerate}

    \item \textbf{High-Conductivity approximation:} This is the conventional and well-studied scenario where all fundamental charged particles are efficiently produced immediately after inflation and get instantaneously thermalised, and the high-temperature background results in a very large electrical conductivity of the universe. In this high-conductivity limit, the rapid response of the charged particles causes the electric field to decay almost instantaneously, effectively removing it from the post-inflationary dynamics. As a result, during reheating, only the magnetic field survives. This can be easily understood from the dynamical equation
    \begin{subequations}\label{eq18}
    \bea 
\mathcal{P}^{c'}_E + 2\sigma \mathcal{P}^{c}_E = -\mathcal{P}^{c'}_B \\
\mathcal{P}^{c''}_B + \sigma \mathcal{P}^{c'}_B+ 2 k^2 \mathcal{P}^c_B   = 2 k^2 \mathcal{P}^c_E.
\eea
\end{subequations}
We assume the background conductivity $\sigma$ to be constant. Before we embark, on few points are in order. In the above evolution equation for the comoving electromagnetic power spectrum during reheating, we assume the gauge coupling function $f(\eta) = 1$ (see Eq.(\ref{eq11})) after the end of inflation $a>\ae$. Further, the equations do not explicitly depend on the scale factor, and that is due to the underlying conformal invariance of the gauge fields. During reheating, there is no coupling term to break the invariance like the one in Eq.(\ref{eq:action_1}). Our goal is to work in the large conductivity limit. 
 It can be seen that in the limit $\sigma \rightarrow \infty$, and large scale $k \rightarrow 0$, the consistent solution of the above set of equations would be 
\begin{eqnarray}
&\mathcal{P}^{c}_E(\eta) \approx 0 \implies |E|^2 \approx 0, \nonumber\\
&\mathcal{P}^{c}_B (\eta) \approx \mB \implies |B|^2 \approx \frac{\mB}{a(\eta)^4}. \nonumber
\end{eqnarray}
Where the integration constant $\mB = \ae^4 \mathcal{P}^{c}_B (\ee) $ is calculated at the time of inflation end. The present-day magnetic field spectrum $\mathcal{P}^{0}_{B}$ with the field strength $B_0$ can therefore be calculated as 
\begin{eqnarray}
a_0^4 \mathcal{P}^0_B(k) \simeq \are^4 \mB \implies B_0 \simeq \frac {\sqrt{\mB}\ae^2}{{a_0^2}} .
\end{eqnarray} 

We restrict our analysis to parameter regimes that avoid both strong coupling and backreaction issues. Since the electric field is no longer present in this case, there is no energy transfer from the electric field to the magnetic field via the Faraday induction mechanism. 
Therefore, the magnetic field evolves adiabatically for those modes that remain super-horizon during reheating.
Therefore, the magnetic energy density redshifts as $ a^{-4} $ due to the expansion of the universe. Consequently, the comoving magnetic field power spectrum at the end of reheating is given approximately by
$
\tilde{\mathcal{P}}_B(k, \eta_{\rm reh}) \simeq \tilde{\mathcal{P}}_B(k, \ee)\,.
$
The present-day magnetic field strength at $ 1\,\mathrm{Mpc} $ scales for a coupling parameter $ n = -2 $ is calculated as
$
B_0(1\,\mathrm{Mpc}^{-1}) \simeq 1.45 \times 10^{-36}\,\mathrm{G}\,.
$
In contrast, for the lower bound of the coupling parameter, $ n \simeq -2.2 $, the present-day magnetic field strength is enhanced to
$
B_0(1\,\mathrm{Mpc}^{-1}) \simeq 1.3 \times 10^{-31}\,\mathrm{G}\,.
$
These estimates suggest that for the coupling parameter $-2.2 < n < 0$, it cannot produce sufficient strength of the magnetic field at $1$ Mpc scale.  
However, for the case of vanishingly small conductivity, we indeed show that sufficient magnetic field strength can be obtained 

    \item \textbf{Low Conductivity approximation:} In this case, we assume that the electrical conductivity of the universe remains small during reheating. This assumption is strongly motivated by the fact that, during reheating, the electrically neutral inflaton field is the dominant energy density until the end of reheating. 
In such a scenario, we are going to discuss how reheating helps us to explain the current observational limit.
They have the same evolution equation defined in Eq.\eqref{eq18}.
As stated earlier, we assume $\sigma \rightarrow 0$ limit.  So, we can identify the solutions of Eqs(\ref{eq18}) as comoving electric and magnetic power spectrum defined as $\mathcal{P}^c_B=a^4\mathcal{P}_B$ and $\mathcal{P}^c_E=a^4\mathcal{P}_E$. And the solution of the Eqs(\ref{eq18}) is governed by
   \begin{subequations}
   \begin{align}\label{eq:pb_sol_re}
    a^4\mathcal{P}_B=\frac{\mB+\mE}{2}+\frac{\mB-\mE}{2}\mbox{ cos}(2k(\eta-\ee)),\\
a^4\mathcal{P}_E=\frac{\mB+\mE}{2}-\frac{\mB-\mE}{2}\mbox{ cos}(2k(\eta-\ee)),
\end{align}
\end{subequations} 
where $(\mB, \mE)$ are the appropriate initial conditions,
\begin{align}\label{eq23}
\mathcal{E}=\frac{\ae^4\pei(\ei)}{f^2(\ei)}=\frac{k^4}{8\pi}\frac{2^{2n+1}\Gamma^2\left(n+\frac{1}{2}\right)}{\pi^2} \left(\frac{k}{\ke}\right)^{-2n} ,
\end{align}
\begin{align}\label{eq24}
\mathcal{B}=\frac{\ae^4\pbi(\ei)}{f^2(\ei)}=\frac{k^4}{8\pi}\frac{\Gamma^2\left(n-\frac{1}{2}\right)2^{2n-1}}{\pi^2}\left(\frac{k}{\ke}\right)^{2-2n}
\end{align}
 In de Sitter space, we can identify it through $\left|-\ei\right|=1/\ke$ where $\kf$ is the highest wavenumber that crossed the horizon at the end of inflation. The above two Eqs. (\ref{eq23}),(\ref{eq24}) are evaluated at the end of inflation for the super-horizon modes approximation. 

During the post-inflationary phase, if there is no further gauge field production, the electromagnetic energy density scales as $\rho_{\rm em} \propto a^{-4}$~\cite{Campanelli:2008kh, Jain:2012jy, Adshead:2016iae, Kahniashvili:2016bkp, Tripathy:2021sfb, Li:2022yqb}. However, it is important to note that the individual evolution of the electric and magnetic field components, $E^2$ and $B^2$, may deviate from each other, as their behavior depends sensitively on the electrical conductivity of the universe during reheating~\cite{Benevides:2018mwx, Kobayashi:2019uqs, Haque:2020bip}. This nontrivial evolution can be attributed to Faraday induction, through which electric energy can be converted into magnetic energy, and vice versa, depending on the field strengths~\cite{Haque:2020bip, Kobayashi:2019uqs}. For example, the case we just discussed in the above Eq.\ref{eq:pb_sol_re}, where the Faraday effect is evident. However, it would be pronounced in large-scale limit $k\eta \rightarrow 0$, when the above equations are approximated as,
   \begin{align}\label{eq:pb_sol_re1}
\mathcal{P}_B = \frac {1}{a^4}(\mB + (\mB -\mE)k^2(\eta-\ee)^2) \simeq \frac {\mB}{a^4} + \frac {\mB-\mE}{H^2 a^6} k^2 ,\\
\mathcal{P}_E=\frac {1}{a^4}(\mE-({\mB-\mE})k^2(\eta-\ee)^2) \simeq \frac {\mE}{a^4} - \frac {\mB-\mE}{H^2 a^6} k^2,
\end{align}
Where, we utilize the relation $\eta -\ee \sim 1/(a H)$. The above equation clearly indicates how the Faraday effect transfers energy from electric to magnetic modes and vice versa with time behavior $\mathcal{P}_B \propto 1/a^6H^2$, which has been discussed earlier in the literature \cite{Kobayashi:2014sga,Haque:2020bip}.

In what follows, we focus on the intermediate regime $ -2.2 < n < 0 $, corresponding to $ -0.4 < \ne < 4 $, where both strong coupling and backreaction problems are absent. In this case, as stated earlier electric field is stronger than the magnetic field at the end of inflation,  $\mE \gg \mB $. The Faraday effect, therefore, leads to an additional enhancement of the magnetic field, and at the end of reheating, one obtains,
\begin{align}
 &   \mathcal{P}_E(k,\ere)\simeq \frac{1}{\are^4} (\mE+(\mB-\mE)k^2\ere^2) \simeq\frac{\mE}{\are^4}  
 \\
 &   \mathcal{P}_B(k,\ere)\simeq \mB-(\mB-\mE)k^2\ere^2 \simeq \frac {\mE k^2}{\are^6 H_{\rm re}^2} .
\end{align}
After the end of reheating, assuming the universe is dominated by the relativistic plasma with infinite conductivity, the electric field quickly decays, and the magnetic field evolves adiabatically till the present day. The present-day magnetic field $B_0$ can be calculated as 
\begin{eqnarray}
a_0^4 \mathcal{P}^0_B(k,\ere) \simeq \frac{\mE k^2}{\are^2 H_{\rm re}^2} \implies B_0 \simeq \frac {\sqrt{\mE} k }{\are H_{\rm re}{a_0^2}}  
\end{eqnarray} 
Here, we have observed that due to the Faraday induction effect, there is significant conversion of electric field energy into magnetic field energy. To highlight this effect, we present some numerical comparisons for two extreme cases of electrical conductivity.
For the zero electrical conductivity (EC) case with coupling parameter $ n = -2.0\,(\ne = 0) $, where the magnetic field energy density scales as $ k^2 $, we find that for a reheating temperature $ \Tre = 10^{-2}\,\mathrm{GeV} $ and equation of state $ \wre = 0 $, the present-day magnetic field strength is $ B_0(1\,\mathrm{Mpc}^{-1}) \simeq 1.86 \times 10^{-27}\,\mathrm{G} $. In contrast, for the same parameters but assuming infinite EC, the strength drops to $ B_0(1\,\mathrm{Mpc}^{-1}) \simeq 8.85 \times 10^{-40}\,\mathrm{G} $.
Similarly, for the same reheating temperature with $ \wre = 1/3 $, we find that the present-day magnetic field strength in the zero EC case is $ B_0(1\,\mathrm{Mpc}^{-1}) \simeq 1.27 \times 10^{-15}\,\mathrm{G} $, while the infinite EC case yields $ B_0(1\,\mathrm{Mpc}^{-1}) \simeq 7.3 \times 10^{-34}\,\mathrm{G} $.
Therefore, it is evident that for a fixed reheating temperature, the zero conductivity scenario leads to a significant enhancement in the magnetic field strength, up to a factor of $ 10^{14}-10^{19} $ compared to the infinite conductivity case, due to efficient electric-to-magnetic energy conversion.

\end{enumerate}

\paragraph{\underline{\bf Constraining the Reheating parameter from the Backreaction at the end of Reheating:}}
Before we embark on the production of gravitational waves, let us derive a generic constraints on the reheating parameter space due to additional production of gauge field. 
As we now consider a general reheating history, it is important to account for the possibility that the equation of state (EoS) parameter during reheating satisfies $ \wre > 1/3 $. In such cases, the background energy density dilutes faster than the electromagnetic (EM) energy density. Specifically, the inflationary energy density evolves as $ \rho_{\phi}(\eta) \propto a^{-3(1 + \wre)} $, whereas the electromagnetic energy density evolves as $ \rho_{\rm em}(\eta) \propto a^{-4} $. This implies that the fractional energy density of the EM field increases during reheating according to
$
\delta\rho_{\rm em}(\eta) \equiv \frac{\rho_{\rm em}(\eta)}{\rho_{\phi}(\eta)} \propto a^{3\wre - 1}\,.
$
Therefore, depending on the reheating dynamics, there may arise a situation in which the EM energy density becomes comparable to or even exceeds the background energy density.

This scenario is particularly constrained by Big Bang Nucleosynthesis (BBN). If the universe is dominated by magnetic field energy at the onset of BBN, it might disrupt the formation of light elements. Since the predictions of BBN are in excellent agreement with observations, we must impose a consistency condition that the total EM energy density at the end of reheating remains subdominant to the background radiation energy density, $\rho_{\rm em}(\eta_{\rm reh}) <\rho_r(\eta_{\rm reh})\,$.
Imposing this condition leads to nontrivial constraints on the reheating dynamics.

These constraints are further modified depending on the post-inflationary conductivity of the universe. Let us first consider the case of high electrical conductivity. In this limit, only the magnetic field survives during reheating. As discussed previously, and within the allowed parameter range $ -2.2 < n < 0 $, the magnetic power spectrum is blue-tilted and evolves adiabatically, as described by Eq.~\eqref{eq:pbi_adibatic}. In this high-conductivity regime, the fractional energy density of the magnetic field evolves as
$
\delta\rho_B(\eta) \equiv \frac{\rho_B(\eta)}{\rho_{\phi}(\eta)} \propto a^{3\wre - 1}\,,
$
similar to the general EM case. Thus, for sufficiently large $ \wre $, the magnetic field can become dynamically significant and potentially dominate the energy budget, leading to tight constraints on the reheating duration and EoS.
\begin{align}\label{eq:delta_B_c}
    \delta\rho_B(\ere)\simeq \l(\frac{\HI}{\Mp}\r)^2\frac{2^{3-\ne}\Gamma^2\l(\frac{3-\ne}{2}\r)(1-(k_*/\ke)^{\ne+2})}{24\pi^3(\ne+2)}\exp[\Nre(3\wre-1)]
\end{align}
\begin{figure}[t]
\begin{center}
\includegraphics[scale=0.43]{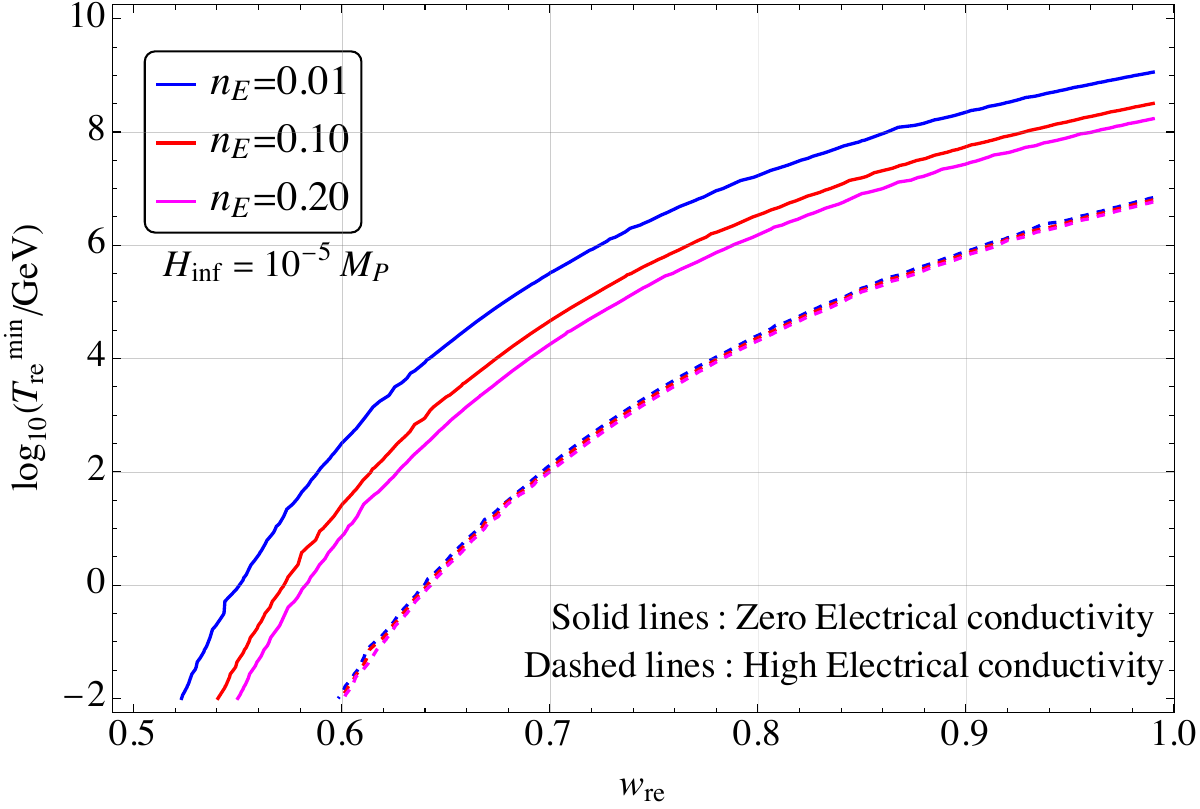}
\caption{In this figure, we present the lowest allowed reheating temperature (in GeV) as a function of the reheating equation of state parameter $ \wre $, over a wide range. The three different colors correspond to three distinct values of the electric spectral index: $ \ne = 0.01 $ (\textbf{blue}), $ \ne = 0.1 $ (\textbf{red}), and $ \ne = 0.2 $ (\textbf{magenta}). Solid lines represent the scenario with zero electrical conductivity (EC), while dashed lines correspond to the high EC limit. Throughout the analysis, we fix the inflationary energy scale at $ H_{\rm inf} \simeq 10^{-5} M_{\rm Pl} $.}
\label{fig:Tre_min}
\end{center}
\end{figure}
On the other hand, in the zero electrical conductivity limit, both electric and magnetic fields are present during the reheating era. We assume that the electric field decays only after reheating completes, when the universe becomes fully dominated by radiation. Therefore, before the end of reheating, both fields contribute to the total electromagnetic energy density. To consistently avoid backreaction in this regime, it is essential to account for the combined influence of both the electric and magnetic fields. Consequently, in the zero conductivity limit, the electromagnetic fractional energy density must be defined by including contributions from both components.
\begin{align}\label{eq:delta_B_zero_c}
    \delta\rho_{em}(\ere)\simeq \l(\frac{\HI}{\Mp}\r)^2\frac{2^{5-\ne}}{24\pi^3}\l[ \frac{\Gamma^2\l(\frac{5-\ne}{2}\r)(1-(k_*/\ke)^{\ne})}{\ne}+\frac{\Gamma^2\l(\frac{3-\ne}{2}\r)(1-(k_*/\ke)^{\ne+2})}{4(\ne+2)}\r]\exp[\Nre(3\wre-1)]
\end{align}
As seen from Eqs.\eqref{eq:delta_B_c} and \eqref{eq:delta_B_zero_c}, the fractional electromagnetic energy density increases with time only in reheating scenarios with $ \wre > 1/3 $. For example, in the high-conductivity limit with $ \wre = 0.5 $, the minimum reheating temperature required to prevent the magnetic field from dominating before the onset of the radiation-dominated era is $ \Tre^{\rm min} \sim 10^{-2}\,\mathrm{GeV} $. In contrast, for the same equation of state but in the zero electrical conductivity limit, the corresponding minimum reheating temperature is significantly higher, $ \Tre^{\rm min} \sim 3 \times 10^{2}\,\mathrm{GeV} $.

In Fig.~\ref{fig:Tre_min}, we present the lowest allowed reheating temperature, denoted by $ \Tre^{\rm min} $ (in GeV), as a function of the reheating equation of state parameter $ \wre $, for three different values of the electric spectral index: $ \ne = 0.01 $, $ 0.1 $, and $ 0.2 $. These parameter choices ensure that the model remains free from both backreaction and strong coupling problems during inflation as well as during reheating. As previously discussed, for $\wre> 1/3$, the background energy density dilutes faster than the electromagnetic (EM) energy density, allowing the latter to potentially dominate at later stages of reheating. The curves in the figure indicate the threshold values of $\Tre$ below which the EM field would dominate over the background energy density before reheating completes.

In this plot, solid lines represent the zero electrical conductivity limit, while dashed lines correspond to the high conductivity limit. As observed, the constraints in the high conductivity scenario are less sensitive to the spectral index $\ne$, since only the magnetic field contributes, and the total magnetic energy density, being strongly blue-tilted, is relatively insensitive to variations in $\ne$. In contrast, in the zero conductivity case, the majority of the EM energy budget is due to the electric field, whose energy density is highly sensitive to $\ne$ for the chosen parameter range.
\begin{figure}[t]
\begin{center}
\includegraphics[scale=0.41]{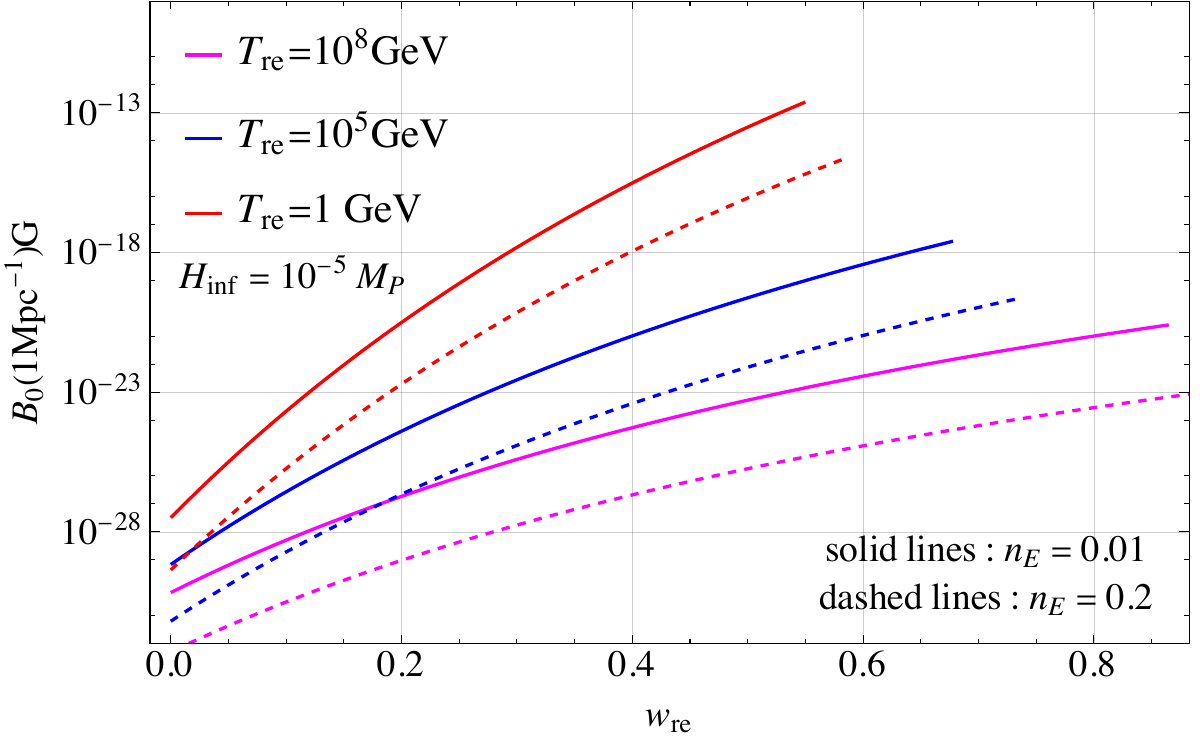}
\includegraphics[scale=0.41]{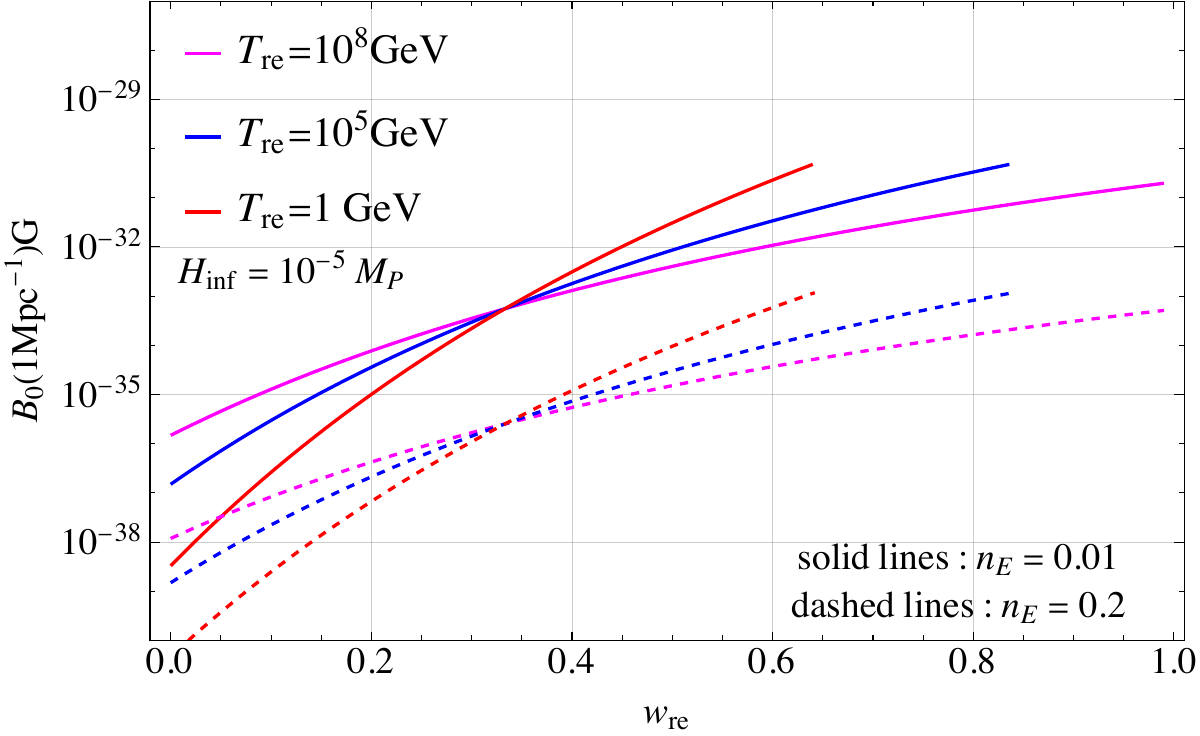}
\caption{In this figure, we show the present-day magnetic field strength evaluated at the scale $ 1\,\mathrm{Mpc}^{-1} $ as a function of the reheating equation of state parameter $ \wre $. The three different colors represent three distinct values of the reheating temperature. Solid lines correspond to $ \ne = 0.01 $, while dashed lines indicate $ \ne = 0.20 $. The left panel illustrates the case of zero electrical conductivity during reheating, whereas the right panel corresponds to the infinite conductivity limit. In both cases, we impose the condition that the total electromagnetic (EM) energy density remains subdominant to the background energy density throughout reheating, thereby ensuring that the universe does not enter a magnetic field-dominated phase before the onset of Big Bang Nucleosynthesis (BBN).}
\label{fig:B0_vs_wre}
\end{center}
\end{figure}

In this work, we neglect the magnetohydrodynamic (MHD) evolution of sub-horizon modes, which are most susceptible to MHD effects. Consequently, for blue-tilted magnetic spectra, the estimates provided here are expected to be significantly altered in more realistic scenarios. However, since a full treatment of MHD dynamics lies beyond the scope of this study, we present our results as indicative bounds.

Using these bounds, we analyze the behavior of the present-day magnetic field under two limiting cases, zero electrical conductivity and infinite electrical conductivity during reheating. In Fig.~\ref{fig:B0_vs_wre}, we plot the present-day magnetic field strength as a function of the reheating equation of state $ \wre $, for three different reheating temperatures: $ \Tre = 1\,\mathrm{GeV},\, 10^5\,\mathrm{GeV},\, \text{and } 10^8\,\mathrm{GeV} $.

In the left panel of Fig.~\ref{fig:B0_vs_wre}, corresponding to the zero conductivity limit, we observe that a significant portion of the electric field energy is converted into magnetic field energy during reheating. As a result, the present-day magnetic field amplitude is considerably enhanced. In contrast, the right panel corresponds to the infinite conductivity limit, where no such conversion takes place. Consequently, the magnetic field strength today is much smaller compared to the zero conductivity case.

We further observe that increasing the equation of state $ \wre $ and decreasing the reheating temperature $ \Tre $ both tend to enhance the present-day magnetic field strength. This enhancement is driven by the dilution of magnetic energy due to cosmic expansion. For larger $ \wre $, the background energy density redshifts more rapidly, allowing the Universe to reach lower reheating temperatures over a shorter timescale. As a result, magnetic fields experience less dilution from expansion, leading to a stronger residual amplitude today.

Moreover, we find that for more blue-tilted electric spectra, the magnetic field strength at large scales decreases. This is because the magnetic spectrum becomes increasingly blue-tilted with growing $ \ne $, suppressing power on large scales. For instance, considering the zero conductivity case with $ \ne \simeq 0.01 $, $ \Tre \simeq 1\,\mathrm{GeV} $, and $ \wre \simeq 0.5 $, the magnetic field today at $ 1\,\mathrm{Mpc}^{-1} $ is $ B_0 \simeq 3.04 \times 10^{-14}\,\mathrm{G} $. In contrast, for the same reheating parameters but in the infinite conductivity case, the present-day magnetic field is significantly weaker, $ B_0 \simeq 3.01 \times 10^{-32}\,\mathrm{G} $.

These results highlight the crucial role of post-inflationary gauge field evolution in determining the final amplitude of the magnetic field. In particular, the electrical conductivity of the Universe during reheating has a significant impact on the resulting magnetic field strength observable today.


\section{Generation of Gravitational Waves:}\label{sec3}

The early universe provides a rich setting for gravitational wave (GW) generation, with primary contributions from quantum fluctuations during inflation and secondary sources such as primordial electromagnetic (EM) fields. Quantum fluctuations in spacetime, amplified during inflation, imprint a stochastic GW background that carries signatures of the inflationary dynamics. However, beyond this primary GW production, secondary sources—particularly EM fields—can significantly enhance the GW spectrum. Magnetic and electric fields, likely generated through the above-mentioned process, can further enrich the GW signal and offer new observational windows into early universe physics. In this section, we mainly explore the secondary contributions from EM fields, highlighting their distinct and complementary roles in shaping the detectable GW landscape.

The GW perturbed Friedmann-Robertson-Walker (FRW) metric takes the form:
\begin{align}\label{eq34}
ds^2 = a^2(\eta)\left(-d\eta^2 + (\delta_{ij} + h_{ij})dx^i dx^j\right),
\end{align}
where `$\eta$' is the conformal time and $h_{ij}$ is the traceless tensor, i.e., $\partial^i h_{ij} = h^i_i = 0$. In Fourier space, the equation of motion for the gravitational wave amplitude `$h$', for either polarization $h^{+}$ or $h^{-}$, becomes \cite{PhysRevD.65.023517, Sorbo:2011rz, Caprini:2014mja, Ito:2016fqp, Sharma:2019jtb, Okano:2020uyr}:
\begin{align}\label{eq:hk_S}
    \hk'' + 2\mH \hk' + k^2 \hk = \mS(\vk, \eta),
\end{align}
where $\mS$ is the source term. Here we consider electromagnetic fields as the source term, which can be written as
\begin{align}
    \mS_{\lambda}(\vk, \eta) = -\frac{f^2(\eta)}{a^2(\eta)} \int \frac{d^3q}{(2\pi)^{3/2}} e^{ij}_{\lambda}(\vk) \left\{ A'_i(\vq, \eta) A'_j(|\vk - \vq|, \eta) - \epsilon^{iab} q_a A_b(\vq, \eta) \epsilon^{icd} (k - q)_c A_d(|\vk - \vq|, \eta) + \dots \right\}.
\end{align}
In the above, we consider only those relevant terms contributing to tensor production. Here $e^{ij}_{\lambda}(\vk)$ is the polarization tensor, and for a non-helical magnetic field, it is convenient to choose a linear polarization basis defined as $e^{\pm}_{ij} = \frac{1}{\sqrt{2}} \left( \hat{e}_1 \times \hat{e}_2 \pm \hat{e}_2 \times \hat{e}_1 \right)_{ij}$, 
which satisfies $\delta_{ij} e^{ij}_{\lambda}(\vk) = \vk_i e^{ij}_{\lambda}(\vk) = 0$ and $e_{il}^{\lambda}(\vk) e_{lj}^{\lambda}(-\vk) = \delta_{ij}$.  Here $\hat{e}_{1}\,\&\, \hat{e}_{2}$ are a set of mutually orthonormal basis vectors of our coordinate system. The inhomogeneous solution of Eq.~(\ref{eq:hk_S}) can then be written as
\begin{align}
    \hk(\eta) = \int_{\eta_i}^{\eta} d\teta~ \mGk(\eta, \teta) \mS(\vk, \teta),
\end{align}
where $\mGk$ is the Green's function of associated operator of Eq.~(\ref{eq:hk_S}).
The GW spectrum is defined as
\begin{align}\label{gen-def-pow}
\langle h^{(\lambda)}_{\vk} h^{(\lambda)}_{\vk'} \rangle = \frac{2\pi^3}{k^3} \mathcal{P}^{(\lambda)}(k, \eta) \delta^{(3)}(\vk + \vk').
\end{align}
It is convenient to express the tensor power spectrum as the sum of two components: one arising from vacuum fluctuations and the other from source terms. These two contributions are uncorrelated, so we write them as:
\begin{align}\label{powe_spl}
\Pt(k, \eta) = \Ptp(k, \eta) + \Pts(k, \eta),
\end{align}
where $\Ptp(k, \eta)$ refers to vacuum production, and $\Pts(k, \eta)$ is due to the EM field. For a general case with a power-law type electromagnetic spectrum $\mathcal{P}_{\rm B/E}(k, \eta) \propto (k\eta)^{\nbe}$, the tensor power spectrum at any time $\eta$ can be written as \cite{Maiti:2024nhv}
\begin{align}\label{eq:pts-gen}
\Pts(k, \eta) &= \frac{2}{\Mp^4} \left[ \int_{\eta_i}^{\eta} d\eta_1 a^2(\eta_1) \Gk(\eta_f, \eta_1) \pbe(k, \eta_1) \right]^2 \times \int_{0}^{\infty} \frac{dq}{q} \int_{-1}^1 d\mu \frac{(q/k)^{\nbe} f(\mu, \beta, \lambda)}{\left[1 + \left( \frac{q}{k} \right)^2 - 2\mu \frac{q}{k} \right]^{\frac{3 - \nbe}{2}}} 
\end{align}
where $f(k, q, \lambda) = (1 + \mu^2)(1 + \beta^2)$, with $\mu = \hat{\bf k} \cdot \hat{\bf q}$ and $\beta = \widehat{\bf k - q} \cdot \hat{\bf k}$ \cite{PhysRevD.69.063006, Zucca:2016iur, Sharma:2019jtb, Maiti:2024nhv}. Here, $\nbe$ is the spectral index for the magnetic or electric field, depending on which field dominates during this era.

\subsection{Generation and Evolution of Primary Gravitational Waves}
Considering a well-known slow-roll inflation scenario, approximated by a de Sitter universe with the scale factor evolving as $a(\eta) = \left(1 - \HI \eta\right)^{-1}$, where $\HI$ represents the Hubble constant during inflation.
 Where the homogeneous solution of Eq.~(\ref{eq:hk_S}), with Bunch-Davies initial conditions, is given by~\cite{Haque:2021dha}
\begin{align}
    \hk(\eta) = \frac{\sqrt{2}}{\Mp} \frac{i \HI}{\sqrt{2k^3}} \left[ 1 - \frac{i k}{\HI a(\eta)} \right] e^{-i k / \HI} e^{i k / (\HI a(\eta))}.
\end{align}
This demonstrates that the amplitude of the tensor fluctuations is directly related to the energy scale of inflation, i.e., $\hk \propto \HI$. At the end of inflation $(\eta = \ee)$, we can define the tensor power spectrum as~\cite{Starobinsky:1979ty, Grishchuk:1974ny, Guzzetti:2016mkm, Haque:2021dha}
\begin{align}\label{ptv-inf}
    \Ptpi(k) = \frac{2}{\pi^2} \left( \frac{H_{\rm I}}{\Mp} \right)^2 \left( 1 + \frac{k^2}{\ke^2} \right),
\end{align}
where $\ke$ is the highest mode that exited the horizon during inflation.

After inflation, the universe undergoes different phases, and the initial fluctuations are modified as they pass through these phases, encoding information about the universe's history. We consider a non-instantaneous reheating scenario, where the scale factor evolves as $a(\eta) \simeq \ae \left( \eta / \ee \right)^{\delta / 2}$, with $\delta = 4 / (1 + 3 \wre)$, where $\wre$ is the average equation of state (EoS) during reheating. Solving the homogeneous part of Eq.~(\ref{eq:hk_S}) during the reheating phase, the tensor fluctuations $\hk$ in the super-horizon approximation $(k < \ke)$ are given by
\begin{align}
    \hk(\eta > \ee) \simeq \frac{\pi^2 x^l}{2 \Gamma(l) \Gamma(1 - l)} \left[ \frac{2 - l}{\Gamma(2 - l)} \left( \frac{k}{\ke} \right)^{2(1 - l)} J_l(x) + \frac{1}{\Gamma(l)} J_{-l}(x) \right] \hk(\ee),
\end{align}
where $x = k \eta$, $l = {3 (\wre - 1)}/{2 (1 + 3 \wre)}$ and $J_l(x)$ is The Bessel functions of the first kind. Here, $\hk(\ee)$ is the tensor fluctuation produced during inflation, defined at the end of inflation, and includes contributions from both quantum fluctuations and source terms, such as EM fields.

After reheating, the universe enters a radiation-dominated phase, where the background energy density scales as $a^{-4}$ due to expansion. During this phase, the scale factor evolves as $a(\eta) \propto \eta$. Solving the homogeneous part of Eq.~(\ref{eq:hk_S}) in this era, the tensor fluctuation $\hk(\eta > \ere)$ takes the form
\begin{align}
    \hk(\eta > \ere) = x^{-1} \left( \mathcal{D}_1 e^{-ix} + \mathcal{D}_2 e^{ix} \right),
\end{align}
where $\mathcal{D}_1$ and $\mathcal{D}_2$ are constants defined by
\begin{subequations}\label{eq:D}
    \begin{align}
        \mathcal{D}_1(\xre) &= \frac{\hkre(\xre) \left( i \xre - 1 \right) - \xre \hkrep(\xre)}{2i} e^{i \xre}, \\
        \mathcal{D}_2(\xre) &= \frac{\xre \hkrep(\xre) + \hkre(\xre) \left( i \xre + 1 \right)}{2i} e^{-i \xre}.
    \end{align}
\end{subequations}
Here $\hkre$ and $\hkrep$ are the defined tensor fluctuations and their 1st derivative at the end of reheating. It contains both the inflationary effects of the primary production and secondary production of the tensor perturbation. Using Eq.~(\ref{eq:D}), we define the tensor power spectrum during the radiation-dominated era  (due to the homogeneous solution of the tensor equation) as 
\begin{align}
    \mPt(k, \eta > \ere) \simeq \frac{1}{2 (k \eta)^2} \left( 1 + \left( \frac{k}{\kre} \right)^2 \right) \mPt(k, \ere).
\end{align}
Here, $\mPt(k, \ere)$ represents the total tensor power spectrum at the end of reheating. For modes that remain outside the horizon at the end of reheating $(k < \kre)$, the tensor power spectrum in the radiation-dominated era scales proportionally to the inflationary tensor power spectrum, i.e., $\mPt(k, \eta > \ere) \propto \mPt(k, \ee)$. However, for modes that re-entered the horizon before reheating ended $(k > \kre)$, the tensor power spectrum is affected by the reheating history, scaling as $\mPt(k > \kre) \propto \xre^{2l + 1} \mPt(k, \ee)$, where $\xre = k \ere$.

\subsection{Generation of Secondary Gravitational Waves}
Electromagnetic (EM) fields in the early Universe provide a compelling secondary source of gravitational waves (GWs), with the potential to enhance the GW spectrum beyond the standard inflationary background. There are three distinct cosmological phases during which EM fields can source GWs: during inflation, during reheating, and in the subsequent radiation-dominated era. In this work, we focus on scenarios where EM fields are generated during inflation (as discussed in Sec.~\ref{sec2}), potentially attaining significant amplitudes depending on the model parameters.

In such models, EM fields serve as key sources for GWs throughout both the inflationary and post-inflationary epochs. However, successful magnetogenesis in these scenarios—avoiding strong coupling and backreaction problems—requires the coupling parameter to lie within the range $ -2.2 < n < 0 $. Within this regime, the electric field typically dominates on large scales, implying that the corresponding tensor perturbations sourced by the EM field are predominantly generated by the electric component.

Moreover, the post-inflationary evolution plays a crucial role in determining the present-day magnetic field strength. Specifically, in the case of vanishing electrical conductivity during reheating, the induction effect leads to significant amplification of large-scale magnetic fields, even for $ n < 0 $ couplings, while still evading strong coupling and backreaction issues. Therefore, in this section, we focus on scenarios that produce large-scale magnetic fields within the allowed parameter range and remain free from theoretical inconsistencies.

To compute the tensor power spectrum sourced by the electric field during inflation, we solve Eq.~\eqref{eq:hk_S} using the Green's function method, following the approach of \cite{Maiti:2024nhv}. In a de Sitter background, the Green’s function is well established and has been extensively studied in the literature (see, for example, \cite{PhysRevD.85.023534}).
\begin{align}
\Gki(\eta,\eta')=\frac{1}{k^3\eta^{'2}}\left[(1+k^2\eta\eta')\sin(k(\eta-\eta'))+k(\eta'-\eta)\cos(k(\eta'-\eta))\right]\Theta(\eta-\eta').
\end{align}
Assuming that at the end of inflation $\eta\rightarrow 0$, the Green's function simplifies, and we can write it as
\begin{align}\label{green_inf}
\Gki(0,\eta')=\frac{1}{k^3\eta^{'2}}\left[-\sin(k\eta')+k\eta \cos(k\eta')\right].
\end{align}
It is evident that the primary contribution to Green's function arises from the superhorizon scale, particularly in the limit $|k\eta'| \ll 1$. In this regime, we can approximate $\Gk(\eta') \approx \eta'/3$. Utilizing these considerations and incorporating all assumptions, we proceed to derive the tensor power spectrum after inflation within the de Sitter inflationary background
\begin{align}\label{pts-inf}
\Ptsi(k,\ee)=2\left(\frac{\HI}{\Mpl}\right)^4\mCe^2(\ne)\left(\frac{k}{\kf}\right)^{2\ne}\mIinf^2~\fne(k),
\end{align}
where $\mIinf\simeq 1/3\ne$ and we also define,
\begin{align}
    \mCe(\ne)=\frac{2^{5-\ne}~\Gamma^2\left(\frac{5-\ne}{2}\right)}{8\pi^3}
\end{align}
and
\begin{align}\label{fnbe}
\fne(k)=\frac{8}{3\ne}\left(1-\left(\frac{\kmin}{k}\right)^{\ne}\right)+\frac{56}{15(2\ne-3)}\left(\left(\frac{\ke}{k}\right)^{2\ne-3}-1\right).
\end{align}
In this scenario, since the dominant contribution arises entirely from the electric field, the tensor power spectrum sourced by the electric field is expected to scale as 
$\Ptsi(k) \propto \left({\HI}/{\Mp}\right)^4\left({k}/{\ke}\right)^{\ne}$, given that the electric field power spectrum behaves as $
\mathcal{P}_E(k) \propto \HI^4\left({k}/{\ke}\right)^{\ne}$. In contrast, the tensor power spectrum originating from the vacuum quantum fluctuations of spacetime follows $\Ptvi(k) \propto \left({\HI}/{\Mp}\right)^2$.

Typically, for blue-tilted electric spectra with $ \ne > 0 $, the vacuum contribution dominates on large (CMB) scales, since the electric field power is concentrated at small scales. However, in the case of $ \ne < 0 $ (corresponding to $ n < -2 $), the electric field becomes more red-tilted, concentrating its energy on large scales. This opens the possibility that, at CMB scales, the secondary tensor modes sourced by the electric field could exceed the primary vacuum contribution.

After inflation, all fundamental particles are produced through the decay of the inflaton field. The nature of gravitational wave (GW) production in the subsequent reheating era depends sensitively on the electrical conductivity of the Universe. As previously discussed, in the limit of negligible conductivity, both electric and magnetic fields are present and evolve freely. In contrast, in the case of large or infinite conductivity, the electric field is quickly suppressed due to the high response of charged particles, leaving only the magnetic field. Additionally, in the zero-conductivity limit, energy transfer from the electric to the magnetic field via the Faraday induction mechanism significantly enhances the magnetic field amplitude. However, such enhancement does not occur in the high-conductivity regime, resulting in much weaker magnetic fields.

Consequently, in the infinite conductivity limit, the secondary GW production sourced by the magnetic field is expected to be suppressed, particularly at large scales. Since we are interested in scenarios that lead to successful magnetogenesis without encountering strong coupling or backreaction issues, we focus on the case where conductivity during reheating is negligible and the coupling parameter lies within the range $ -2.2 < n < 0 $. In this regime, the dominant post-inflationary contribution to the stochastic gravitational wave background (SGWB) arises from the electric field.

During the reheating phase, the corresponding Green's function can be expressed as~\cite{Maiti:2024nhv}
\begin{align}
    \mGk(x,x_1) 
= \theta(x-x_1)\frac{\pi x^{l} x_1^{1-l}}{2k\mathrm{sin}(l\pi)}
\l[J_l(x)J_{-l}(x_1)-J_{-l}(x)J_l(x_1)\r]
\end{align} 
Since the tensor power spectrum is sourced by the electric field, the tensor spectrum generated during reheating due to the electromagnetic field, evaluated at the end of reheating, can be generally expressed as
\begin{align}\label{eq:ptres}
    \Ptsre=\frac{2\HI^4}{\Mp^4}\left(\frac{k}{\ke}\right)^{2(\delta-2)}\mCe^2(\ne)\left(\frac{k}{\kf}\right)^{2\ne}\Cmre^2(\xre,\xe)\fne(k),
\end{align}
where $\fne$ is defined in Eq. (\ref{fnbe}) and $\Cmre(\xre,\xe)$ is defined as \cite{Maiti:2024nhv}
\begin{align}
    \Cmre(\xre,\xe)=\int_{\xe}^{\xre}dx_1\,x_1^{-\delta}\mGk(\xre,x_1),
\end{align}
Upon analyzing the spectrum, we find that the presence of a non-instantaneous reheating phase introduces an overall factor of the form 
$\left( {\kre}/{\kappa_{\rm end}} \right)^{2(\delta - 2)}$,
where $ \delta ={4}/{(1 + 3\wre)} $. This indicates that the amplitude of the spectrum can either be enhanced or suppressed depending on the equation of state during reheating: it is enhanced for $ \wre > 1/3 $ and suppressed for $ \wre < 1/3 $.

Moreover, we observe that for modes that remain outside the horizon during reheating ($ k < \kre $), the secondary tensor spectrum behaves as
$\mathcal{P}^{\rm re}_{\rm s}(k \ll \kre, \eta_{\rm re}) \propto k^{2\ne}$.
In contrast, for modes that re-enter the horizon during reheating ($ k \gg \kre $), the spectrum scales as
$\mathcal{P}^{\rm re}_{\rm s}(k \gg \kre, \eta_{\rm re}) \propto k^{2\ne - |\nw| - 2},$ where $ \nw ={2(1 - 3\wre)}/{(1 + 3\wre)} $.

For this specific choice of coupling, there are no significant additional contributions to the gravitational wave spectrum from the radiation-dominated era. This is because, following reheating, it is generally assumed that all fundamental particles are produced and the Universe becomes thermalized, resulting in a highly conductive plasma. As a consequence, the electric field component is rapidly suppressed. However, the magnetic field survives and evolves adiabatically, with its energy density decreasing solely due to the expansion of the background.

Therefore, the tensor power spectrum sourced by the magnetic field after reheating, evaluated at the time of neutrino decoupling, is governed by the expression~\cite{Maiti:2024nhv}.
\begin{align}
\Ptrs(k,\eta_\nu)=\frac{2\HI^4}{k^2\eta^2_\nu\Mp^4}\left(\frac{\kre}{\kf}\right)^{2(\delta-2)}\mCb^2(\nb)\left(\frac{k}{\kf}\right)^{2\nb}{\mIra}^2(x_\nu,\xre)\Fnb(k)\label{eq:Ptsra}
\end{align}
We recall that $\nb=\ne+2$ is the magnetic spectral index for $n<0$ magnetogenesis scenarios. Here, $\mIra(x_\nu,\xre)$ is defined as~\cite{Maiti:2024nhv}
\begin{align}
{\mIra(x_\nu,\xre)=\int_{\xre}^{x_\nu}dx_1\frac{\sin(x_1 - x_\nu)}{x_1} }.
\end{align}
As for the low-frequency regions, we found that $\mIra=\gamma_1$, where $\gamma_1\sim 0.5$ at the epoch of neutrino decoupling.

In this context, we observe that the tensor power spectrum generated during this phase includes an overall enhancement factor. This enhancement originates from the dynamics of the reheating phase: while the background energy density dilutes as $ a^{-3(1+\wre)} $, the magnetic field energy density redshifts as $ a^{-4} $. As a result, depending on the equation of state (EoS), the gravitational wave contribution from magnetic fields can either be enhanced (for $ \wre > 1/3 $) or suppressed (for $ \wre < 1/3 $).

Specifically, within the framework of this magnetogenesis model, a significant contribution to the tensor spectrum in the frequency range $ f_\nu < f <\fre $ arises only when $ \wre > 1/3 $ and the coupling parameter satisfies $ n > 1/2 $. Here, $ n $ refers to the coupling constant associated with the magnetogenesis scenario discussed in Sec.~\ref{sec2}.

\subsection{Calculation of the Dimensionless GWs energy density at present time:}
During reheating and the radiation-dominated epoch, perturbations re-enter the Hubble radius, generating a stochastic gravitational wave (GW) background. This GW signal, assumed homogeneous, isotropic, and Gaussian, reflects the properties of the FLRW universe. The tensor power spectrum, $\Pt(k, \eta)$, a dimensionless quantity dependent on the comoving wavenumber $ k $ and conformal time $ \eta $, characterizes this background.

Due to the weak gravitational interaction with matter, GWs decouple on Planck scales, allowing us to neglect matter interactions and self-interactions as sub-Hubble GWs propagate freely post-production. The GW energy density, which scales as $\rho_{\text{GW}} \propto a^{-4}$, can be normalized by the total energy density at production time, $\rho_c(\eta)$. Presently, this density parameter is
\begin{align}
    \ogw(k, \eta) = \frac{\rho_{\text{GW}}(k, \eta)}{\rho_c(\eta)} = \frac{1}{12} \frac{k^2 \Pt(k, \eta)}{a^2(\eta) H^2(\eta)}
\end{align}
where $\rho_c(\eta) = 3 H^2(\eta) \Mpl^2$, with the reduced Planck mass $\Mpl \approx 2.43 \times 10^{18} \, \text{GeV}$.

The GW energy density follows the radiation scaling behavior, making modes within the Hubble radius near radiation-matter equality particularly relevant. The present-day energy density parameter $\ogw(k) h^2$ is then given by
\begin{align}
    \ogw(k) h^2 \simeq \left(\frac{g_{*,0}}{g_{*,\text{re}}}\right)^{1/3} \Omega_R h^2 \ogw(k, \eta),
\end{align}
where $\Omega_R h^2 = 4.3 \times 10^{-5}$, $g_{*,\rm re} = 106.7$, and $g_{*,0} = 3.35$ denote relativistic degrees of freedom at the end of reheating and and today, respectively. 

For simplicity, we decompose the total dimensionless energy density associated with gravitational waves (GWs) into two distinct components, denoted as $\ogw(k) = \ogwp(k) + \ogws(k)$. Specifically, $\ogwp(k)$ represents contributions originating from quantum vacuum fluctuations, while $\ogws(k)$ encompasses the contributions attributed to electromagnetic fields. To comprehensively examine the spectral characteristics of GWs across the entire observable frequency spectrum, we separately discuss two distinct scenarios based on the reheating dynamics, namely, $\wphi<1/3$ and $\wphi>1/3$.

\subsubsection{Spectral Analysis of Primary Gravitational Waves (PGWs)}

Primary Gravitational Waves (PGWs) refer exclusively to the contributions arising from quantum fluctuations during inflation. To analyze the effects of post-inflationary dynamics on the gravitational wave (GW) spectrum, we consider two distinct regimes: (a) super-horizon scales, $ k < \kre $, where modes remain outside the horizon before the end of reheating, and (b) sub-horizon scales, $ \kre < k < \ke $, where modes re-enter the horizon before the end of reheating. While inflation alone typically generates a scale-invariant tensor power spectrum, these regimes reveal distinct spectral behaviors. The present-day GW spectrum, produced by inflation, can be expressed as \cite{Maiti:2024nhv, Chakraborty:2024rgl}
\begin{align}
    \ogwp(k)\simeq  1.12\cdot 10^{-17} \l(\frac{\Omega_Rh^2}{4.3\cdot 10^{-5}}\r)\l(\frac{\HI}{10^{-5}\Mp}\r)^2\times\l\{\begin{matrix}
        1 & &~~ k<\kre\\
        (k/\kre)^{-\nw} & & ~~ \kre<k<\ke ,
    \end{matrix}\r.\label{eq:ogw_pri}
\end{align}
From this expression, we observe that, for super-horizon modes ($ k < \kre $), the spectrum remains scale-invariant, consistent with the inflationary tensor power spectrum. However, for modes that re-entered the horizon before reheating completed ($ k > \kre $), the spectrum is no longer scale-invariant. Specifically, for reheating scenarios with an equation of state $ \wre > 1/3 $ (stiff-fluid-like), the spectrum has a blue tilt, while for $ \wre < 1/3 $ it has a red tilt. This difference arises because, in $ \wre < 1/3 $ scenarios, GW modes that remain inside the horizon for longer durations experience greater dilution, as the background energy density decays more slowly compared to the GW energy density. Conversely, for $ \wre > 1/3 $ scenarios, modes are amplified by spending longer periods within the horizon, due to the faster decay of background energy density relative to GW energy density.

\subsubsection{Spectral Analysis of Secondary Gravitational Waves (SGWs)}
In analyzing the spectrum of secondary gravitational waves (SGWs) over a broad frequency range, we concentrate on contributions arising from inflationary magnetogenesis, while neglecting the primary gravitational waves sourced by vacuum fluctuations. Two limiting scenarios are considered based on the conductivity of the Universe during reheating. In the first case, we assume zero electrical conductivity, where both electric and magnetic fields are present throughout the reheating phase. In the second scenario, we consider infinite electrical conductivity, in which only the magnetic field survives and evolves adiabatically after inflation.

\paragraph{\underline{\bf Zero Electrical Conductivity during the Reheating Era:}}

Under the assumption of negligible electrical conductivity during reheating, both electric and magnetic fields persist during this epoch. In the parameter range that avoids both strong coupling and backreaction issues, namely $ -0.4 < \ne < 4 $, the electric field typically dominates. Consequently, the SGW spectrum is predominantly sourced by the electric field during inflation and reheating. Although the magnetic field can contribute to SGW production in the subsequent radiation-dominated era (up to neutrino decoupling), the contribution from the reheating phase is generally dominant.

The shape of the resulting GW spectrum is sensitive to the background evolution during reheating. Specifically, for reheating scenarios with an equation of state parameter $ \wre < 1/3 $, the present-day SGW spectrum sourced by the electromagnetic field is
\begin{align}
    \ogws(k) h^2 \simeq & \, 2.26 \times 10^{-26} \left( \frac{\Omega_R h^2}{4.3 \times 10^{-5}} \right) \left( \frac{\HI}{10^{-5} \Mp} \right)^4 \left( \frac{\kre}{\kf} \right)^{2 \ne} \mathcal{I}_{\rm inf}^2 \mathcal{C}_{\beta}^2 (\ne) \, \mathcal{F}_{\beta} (k) \nonumber \\
    & \times \begin{cases}
        \left( \frac{k}{\kre} \right)^{2 \ne}, & k_{*} < k < \kre, \\
        \left( \frac{k}{\kre} \right)^{2 \ne - |\nw|}, & \kre < k < \ke .
    \end{cases}\label{eq:gws_1}
\end{align}
whereas for the reheating scenarios with $\wre>1/3$, the spectral behaviour due to EM field is
\begin{align}\label{eq:sgwpiE}
  \ogwspi(k) h^2 \simeq & \, 2.26 \times 10^{-26} \left( \frac{\Omega_R h^2}{4.3 \times 10^{-5}} \right) \left( \frac{\HI}{10^{-5} \Mp} \right)^4 \left( \frac{\kre}{\kf} \right)^{2 (\nw + \ne)} \mathcal{C}_{\beta}^2 (\ne) \, \mathcal{F}_{\ne} (k) \nonumber \\
  & \times \begin{cases}
      \mathcal{A}_1 \left( \frac{k}{\kre} \right)^{2 \ne}, & k_{*} < k < \kre, \\
      \mathcal{A}_2 \left( \frac{k}{\kre} \right)^{2 \ne - |\nw|}, & k > \kre
  \end{cases}
\end{align}
Here, we clearly observe an additional enhancement factor associated with reheating scenarios where $ \wre > 1/3 $, which arises from the specific evolution of the background dynamics during the reheating phase. This enhancement originates from the differential dilution rates between the background energy density and the electromagnetic (EM) field energy density. For $ \wre > 1/3 $, the background energy density redshifts more rapidly than the EM field, leading to a relative amplification of the SGW signal. The coefficients $ \mA_1 $ and $ \mA_2 $, which characterize this enhancement, are provided in Appendix~\ref{Appendix}.

\paragraph{\underline{\bf Infinite Electrical Conductivity During the Reheating Era:}}
In scenarios with infinite electrical conductivity during reheating, electric fields are rapidly dissipated, implying that any contributions from electric fields to the secondary gravitational wave (SGW) background arise solely from their inflationary production. In contrast, magnetic fields can persist and contribute during both the reheating and radiation-dominated eras. However, as demonstrated, even for reheating scenarios with $ \wre > 1/3 $, the magnetic field amplitude remains relatively weak, rendering the induced SGW signal subdominant compared to the primary gravitational waves generated during inflation.

Therefore, for reheating scenarios with $ \wre < 1/3 $, the SGW spectrum retains the same spectral behavior as described in Eq.~\eqref{eq:gws_1}. Conversely, for $ \wre > 1/3 $, the SGW spectrum is no longer dominated by the magnetic field alone and may receive non-negligible contributions from the residual electric field produced during inflation.
\begin{align}
    \ogws(k) h^2 \simeq & \, 2.26 \times 10^{-26} \left( \frac{\Omega_R h^2}{4.3 \times 10^{-5}} \right) \left( \frac{\HI}{10^{-5} \Mp} \right)^4 \left( \frac{\kre}{\kf} \right)^{2 \nbe} \mathcal{I}_{\rm inf}^2 \mathcal{C}_{\beta}^2 (\nbe) \, \mathcal{F}_{\beta} (k) \nonumber \\
    & \times \begin{cases}
        \left( \frac{k}{\kre} \right)^{2 \ne}, & k_{*} < k < \kre, \\
        \left( \frac{k}{\kre} \right)^{2 \ne +|\nw|}, & \kre < k < \ke .
    \end{cases}\label{eq:gws_3}
\end{align}
To achieve a successful magnetogenesis scenario without encountering strong coupling or backreaction issues, the zero electrical conductivity limit during reheating is particularly favorable within the coupling range $-2.2 < n < 0$ (equivalently, $-0.4 < \ne <4 $). In particular, for the subclass with $ n < -2.0 $ (i.e., $ \ne < 0 $), the electric field produced during inflation is red-tilted. This red tilt enhances the power at large scales and can significantly impact the tensor power spectrum, including at cosmic microwave background (CMB) scales, for both $\wre < 1/3$ and $\wre > 1/3$ reheating scenarios.

Most notably, these scenarios can generate a sizable tensor-to-scalar ratio at CMB scales, potentially exceeding the current observational upper bound $r_{0.05} \leq 0.036$. In the following discussion, we explore how this observational constraint can be used to place stringent bounds on both the reheating dynamics and the magnetogenesis model parameter $n$.

 \begin{figure}[t]  
\begin{center}
 \includegraphics[scale=0.4]{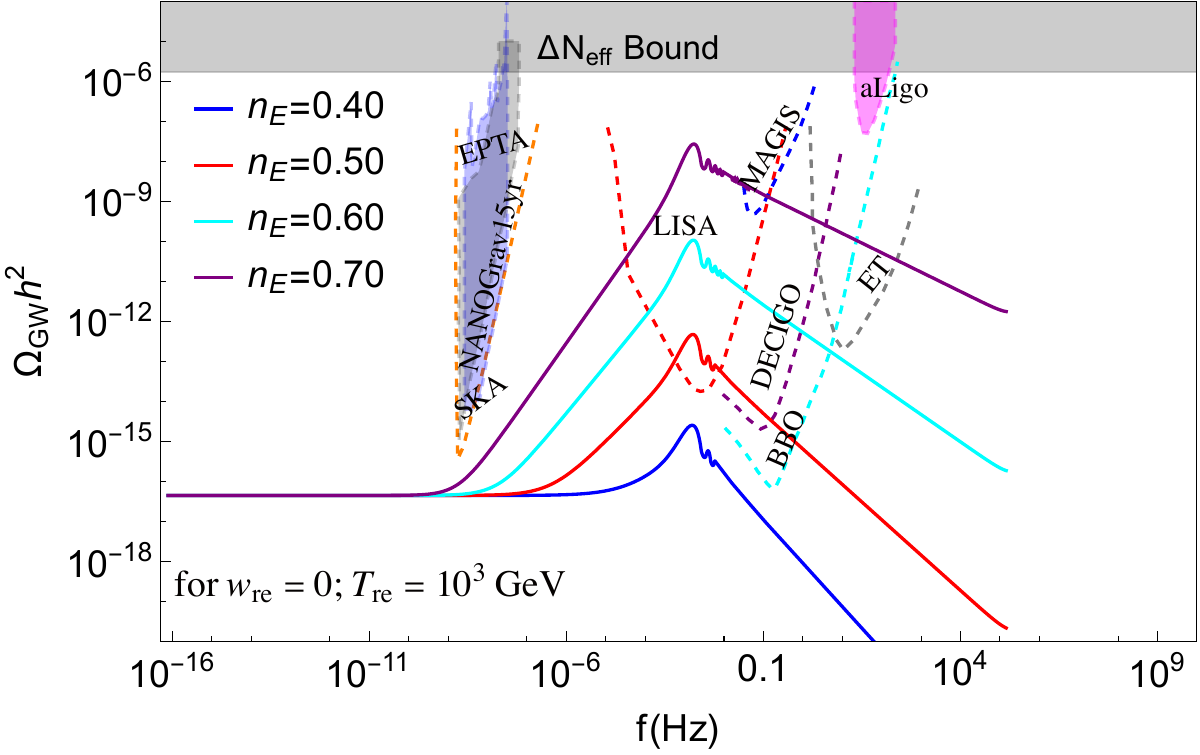}
 \includegraphics[scale=0.4]{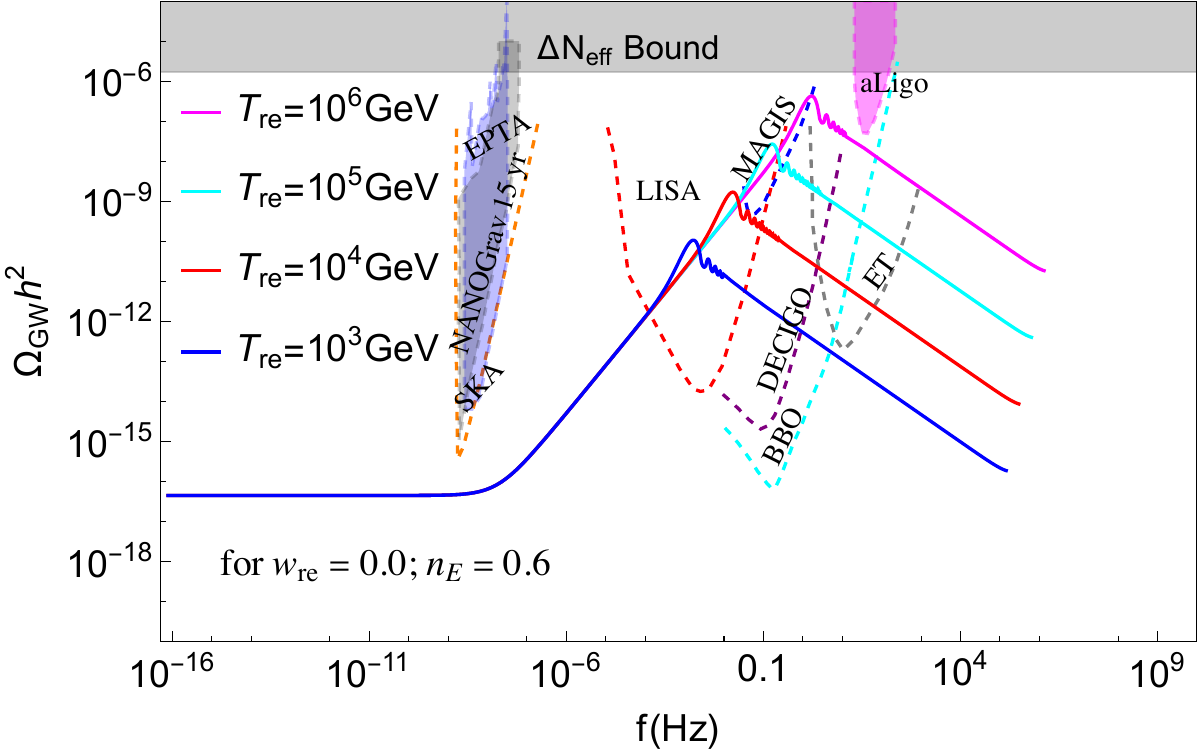}
 \caption{In this figure, we plot $\ogwh$ versus $f$ (in Hz) for $ \wre = 0 $ for two different scenarios. In the left panel, we show how the present-day GWs strength depends on the magnetogenesis parameter $\ne(=4-2|n|)$ for a fixed reheating temperature $\Tre=10^3\,\Gev$. Here, four different colors indicate four different values os the electric spectral index. Whereas in the right panel, we have shown the dependency of the reheating temperature $\Tre$, indicated through four different colors with a fixed electric spectral index $\ne=0.6$. }
 \label{fig:gw_1}
\end{center}
\end{figure}

\begin{figure}[h]  
\begin{center}
 \includegraphics[scale=0.4]{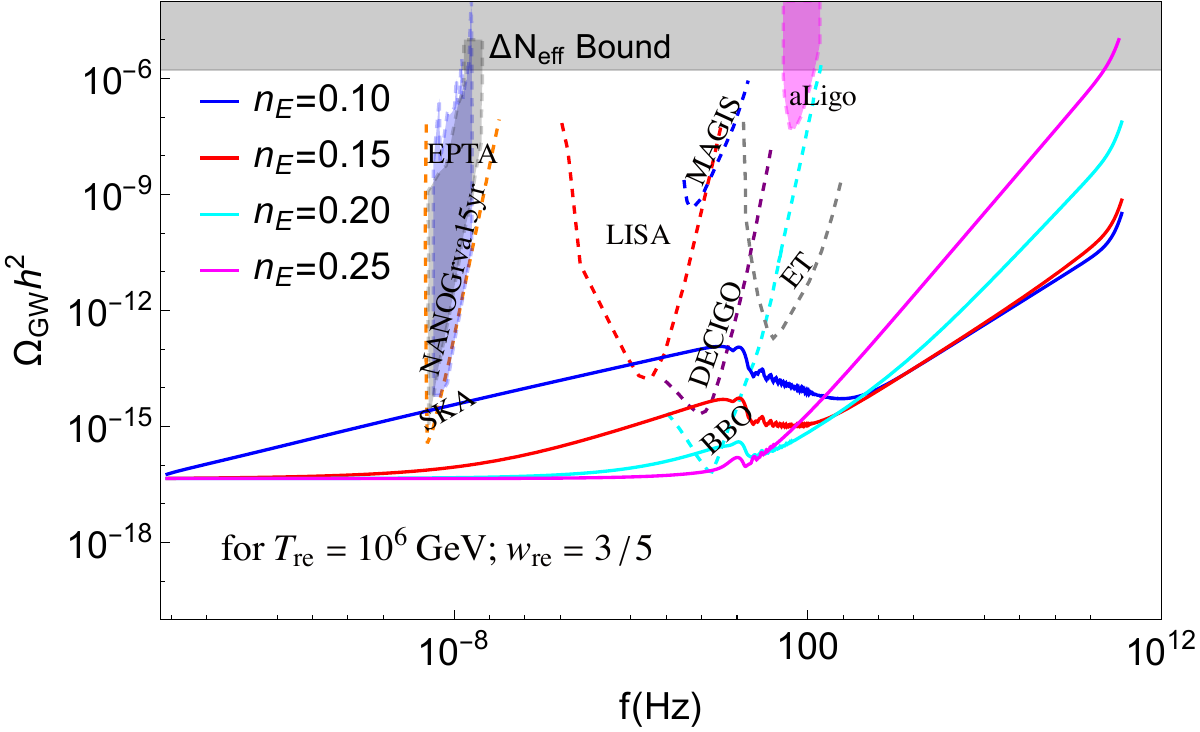}
 \includegraphics[scale=0.4]{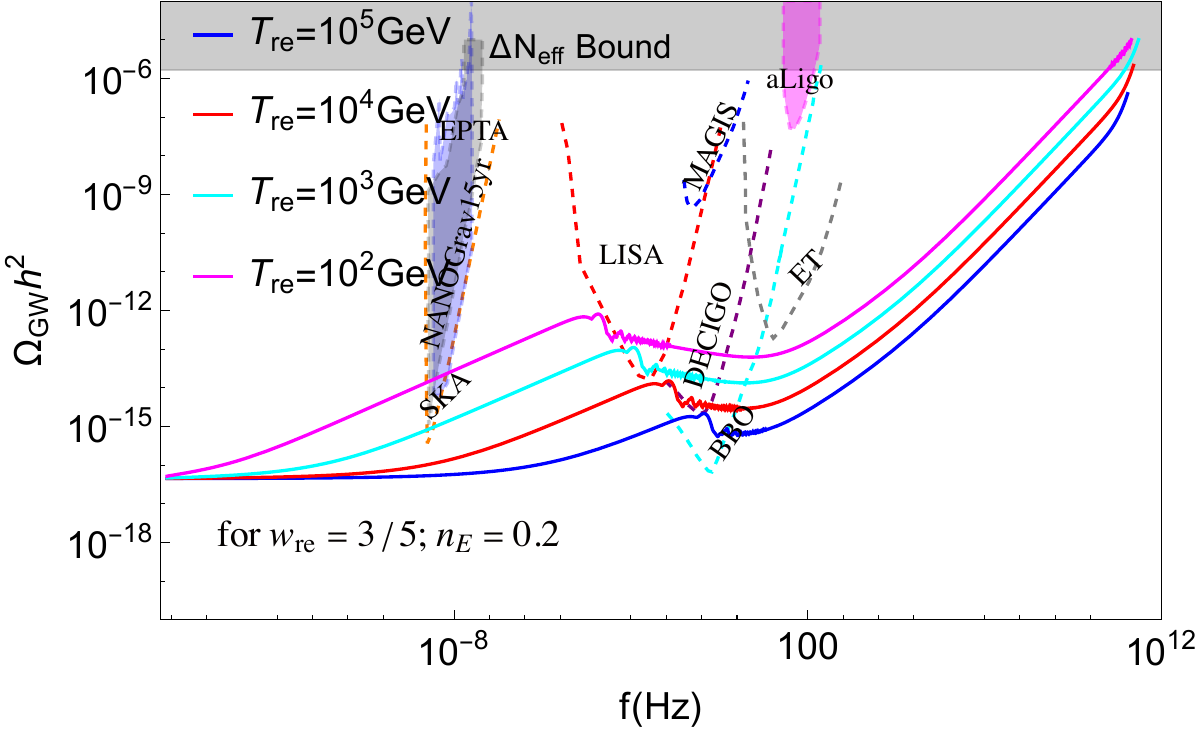}
 \caption{In the figures above, we have plotted $ \ogwh$  as a function of frequency $f$(in Hz) for the coupling $ n < -1/2 $, assuming zero electrical conductivity during the reheating era for $\wre=3/5$ (stiff like fluid during reheating). In the left plot, we have shown the dependency of the GW spectrum as a function of the electrical spectral index $\ne$ defined via four different colors for a fixed reheating temperature $\Tre=10^6\,\ GeV$. In the right panel, we have shown how the present-day GW strength and spectral shape are dependent on the reheating temperature, denoted by four different colors for a fixed electric al spectral index $\ne=0.2$.}
 \label{fig:gw_3}
\end{center}
\end{figure}

\section{Results and Discussion:}\label{sec4}
In Fig.~\ref{fig:gw_1}, we present the present-day gravitational wave (GW) energy spectrum, $\ogwh(f)$, as a function of the observed frequency $f\,(\mathrm{Hz})$, for two distinct reheating scenarios.
In the left panel, we illustrate the impact of the electric spectral index $\ne$ on the GW spectrum, keeping the equation of state fixed at $\wre = 0$ and the reheating temperature at $\Tre = 10^3\,\mathrm{GeV}$. As expected, at very large wavelengths (small $f$), the GW spectrum remains nearly scale-invariant, $\ogwh(f \ll \fre) \propto f^0$, due to the dominance of primary gravitational waves. This behavior is dictated by the scale-invariant nature of tensor modes produced from quantum fluctuations during inflation. 
We consider values $\ne = 0.4,\, 0.5,\, 0.6,$ and $0.7$, corresponding to strongly blue-tilted electric spectra. In such cases, most of the electromagnetic energy is concentrated at high frequencies, resulting in significant GW production in that regime. Consequently, in the intermediate frequency range $f_{\mathrm{cmb}} \ll f < \fre$, the GW spectrum acquires a blue tilt, while at higher frequencies $f > \fre$, the spectrum becomes red-tilted. The spectral slope in this regime can be parametrized as $\ogwh(f > f_{\mathrm{re}}) \propto f^{2\ne - |\nw|}$, where $\nw = 2(1 - 3\wre)/(1 + 3\wre)$.
In the right panel of Fig.~\ref{fig:gw_1}, we demonstrate how the reheating temperature $\Tre$ affects the GW spectrum, fixing $\ne = 0.6$ and $\wre = 0$. In both cases, we observe that for modes that enter the horizon before the end of reheating ($f > \fre$), the spectrum exhibits a red-tilted behavior, driven by the differential dilution rates of the background energy density and the electric field during reheating.

In Fig.~\ref{fig:gw_3}, we present the present-day gravitational wave (GW) spectrum $\ogwh(f)$ as a function of the observable frequency $f$ (in Hz) for reheating scenarios with an equation of state $\wre > 1/3$. For such scenarios, the spectrum typically exhibits more spectral features compared to the $\wre < 1/3$ case, owing to the more complex interplay between primary and secondary GW sources.

The left panel illustrates the dependence of the GW spectrum on the electric spectral index $\ne$, fixing $\wre = 3/5$ and the reheating temperature $\Tre = 10^6\,\mathrm{GeV}$. Depending on the value of $\ne$, we observe a variety of spectral structures: in some cases, the spectrum exhibits three distinct breaks, in others two, and occasionally only one. This rich spectral behavior arises from the varying dominance of the different contributions to the GW background.

For instance, considering $\ne = 0.15$ with $\wre = 3/5$ and $\Tre = 10^6\,\mathrm{GeV}$, the spectrum near the CMB scale appears scale-invariant due to the dominance of primary GWs (PGWs) over secondary GWs (SGWs). At intermediate frequencies $f < \fre$, the spectrum transitions to a blue-tilted behavior $\ogwh(f) \propto f^{2\ne}$, as SGWs sourced by the blue-tilted electric field become dominant. In the range $\fre \ll f \ll f_{\mathrm{end}}$, the spectrum shows a red-tilted behavior due to the suppression of SGWs by reheating dynamics. Finally, at high frequencies $f \gg \fre$, the spectrum becomes blue-tilted again, following $\ogwh(f) \propto f^{|\nw|}$, where PGWs regain dominance.

The right panel of Fig.~\ref{fig:gw_3} shows the impact of the reheating duration on the present-day GW amplitude. For a fixed $\wre = 3/5$, increasing the reheating temperature leads to a gradual suppression of the overall GW amplitude. This behavior is consistent with the scaling $(\ke / \kre)^{2|\nw|}$, which parametrizes the enhancement due to electromagnetic field contributions during reheating for $\wre > 1/3$.

\paragraph{\underline{\bf Constraining model parameters through the Tensor-to-Scalar Ratio $ r $:}}
This magnetogenesis model generates significant electric and magnetic fields capable of producing gravitational waves (GWs) strong enough to pass various sensitivity thresholds and create notable tensor fluctuations at CMB scales. For $\wre > 1/3$ reheating scenarios, the reheating dynamics introduce an overall enhancement factor, $(\ke/\kre)^{|\nw|}$, related to the duration of the reheating phase. However, constraints on tensor fluctuations at CMB scales, quantified by the tensor-to-scalar ratio $ r_{0.05} $, place a limit on GW strength at the pivot scale: $\ogwh \leq 1.14 \times 10^{-7} r_{0.05} A_s$. Here, $ A_s \simeq 2.1 \times 10^{-9} $ is the current bound on the scalar amplitude at the pivot scale.

Using the limit $ r_{0.05} \leq 0.036 $, we find that the coupling parameter $ n $ for magnetogenesis should satisfy $ -2.1 \leq n \leq 0$. This bound holds universally, even for instantaneous reheating. For prolonged reheating scenarios with $\wre > 1/3$, these constraints become more robust, limiting both the reheating dynamics and the magnetogenesis model. Thus, utilizing the upper bound $ r_{0.05} \leq 0.036 $ at the pivot scale, we derive the following constraint,
\begin{align}\label{eq:r-bound}
1 \geq \frac{182~rA_s\mathcal{A}_1}{(3-2\ne)}\left(\frac{\kf}{\kre}\right)^{2|\nw|}\left(\frac{k_{*}}{\kf}\right)^{2\ne}\mCe^2(\ne).
\end{align}
Using this constraint, we set bounds on the reheating temperature for specific values of the equation of state (EoS) across five discrete values of the magnetic spectral index $\ne = 0.01, 0.1, 0.2, 0.3, \text{ and } 0.4$, as shown in Table (\ref{tab:tre_r}).
\begin{table}[t]
    \begin{tabular}{|c|c|c|c|c|c|}
    \hline
        $\wre$ & $\ne=0.01$& $\ne=0.1$ & $\ne=0.2$ &$\ne=0.3$ & $\ne=0.4$ \\
        \cline{2-6}
        & $\Tre$ (GeV) & $\Tre$ (Ge)V & $\Tre$ (GeV)& $\Tre$ (GeV)& $\Tre$ (GeV)  \\
        \cline{1-6}
        0.4 & $1.77\times 10^{9}$ & $-$ & $-$ & $-$ & $-$ \\
        \hline
        0.5 & $5.65\times 10^{12}$ & $1.28\times 10^7$ & $2.89$ & $-$ & $-$\\
        \hline
        0.6 & $8.2\times 10^{13}$ & $1.46\times 10^{10}$ & $5.41\times 10^5$ & $-$ & $-$\\
        \hline
        0.7 & $3.59\times 10^{14}$ & $4.86\times 10^{11}$ & $1.92\times 10^8$ & $4.82\times 10^4$& $7.91$\\
        \hline
    \end{tabular}
    \caption{In the table above, we've listed the minimum reheating temperature for various equations of states (denoted as $\wre$), considering five different values of the electric spectral index (denoted as $\ne$).}
    \label{tab:tre_r}
\end{table}
In our analysis, we treat the reheating equation-of-state parameter $ \wre $ as a free parameter, allowing it to take arbitrary values. However, in specific inflationary models—for example, those with power-law potentials of the form $ V(\phi) \propto \phi^{2m} $, $ \wre $ can be uniquely determined based on the dynamics of the scalar field. In this work, we adopt a generalized approach by leaving $ \wre $ unconstrained.
As an illustration, to avoid the overproduction of tensor perturbations at CMB scales, consider the case where the magnetic spectral index is $ \ne = 0.01 $ (nearly scale-invariant). For a reheating equation-of-state $ \wre = 0.4 $, the lower bound on the reheating temperature is found to be $ \Tre^{\rm min} \simeq 1.77 \times 10^9\,\mathrm{GeV} $. If the reheating temperature falls below this value, electromagnetic fields generated via this mechanism can produce excessive tensor fluctuations at large scales, violating current observational constraints on the tensor-to-scalar ratio.
We summarize the lower bounds on reheating temperatures for various parameter combinations in Table~(\ref{tab:tre_r}).

\paragraph{\underline{\bf Constraining Reheating and Magnetogenesis Parameters through $\Delta N_{\text{eff}}$ Bound:}}
During the epoch of decoupling, the CMB is influenced by the total radiation energy density. Any excess energy that behaves as radiation-like fields may impact the CMB \cite{PhysRevD.85.123002}. In our case, the primordial GWs carry significant energy, and since GWs evolve as $a^{-4}$ like radiation, frequencies $ f > 10^{-15} $ Hz can be treated as extra radiation energy that may affect the CMB (see related discussions in \cite{Caprini:2018mtu, clarke2020constraints}). This excess radiation can be quantified by the additional relativistic degrees of freedom at decoupling, expressed as $\Delta N_{\text{eff}}$, indicating the extra neutrino species predicted beyond the Standard Model. Using this, we derive the following constraint \cite{Caprini:2018mtu, Maiti:2024nhv},
\begin{align}\label{dneff_eq1}
\Delta N_{\text{eff}} \geq \frac{1}{\Omega_{\gamma}h^2} \frac{8}{7}\left(\frac{11}{4}\right)^{4/3} \int_{k_0}^{k_f}\frac{dk}{k}\ogwh(k).
\end{align}
Assuming the present-day photon density parameter as $\Omega_{\gamma}h^2 \simeq 2.48 \times 10^{-5}$, the combined latest Planck-2018 and Baryon Acoustic Oscillation (BAO) data predict $\Delta N_{\text{eff}} \simeq 0.284$ (within a $2\sigma$ range) \cite{Planck:2018vyg}, setting an upper bound on primordial GWs, $\ogwh < 1.67 \times 10^{-6}$ \cite{Clarke:2020bil}.

At high frequencies, the GW spectrum behaves as $\ogwh \propto f^{2\ne}$ for $f < f_{\rm re}$, while for modes well within the horizon $(f \gg f_{\rm re})$, it follows $\ogwh \propto f^{2\ne - \nw}$. Therefore, if $2\ne > \nw$, the spectrum has a red tilt for $f > \fre$; otherwise, it has a blue tilt. Using this bound, we find that for reheating scenarios with $\wre = 0$, the maximum allowed magnetic spectral index consistent with the $\Delta N_{\text{eff}}$ bound is $\ne \leq 1.0$. This constraint also restricts the reheating temperature for a fixed magnetic spectral index and equation of state (EoS). For instance, with $\wre=0$ and $\ne=1.0$, the reheating temperature should be below $\Tre < 10^3$ GeV. For $\ne=0.9$ and $\wre =0$, a more relaxed bound is obtained, $\Tre\leq10^7$ GeV.

For higher EoS, $\wre > 1/3$, at high frequencies with $\ne> 0.3$, the GW spectrum follows $\ogwh \propto f^{2\ne + |\nw|}$. In this regime, secondary GW production during inflation is more substantial compared to late-time production, and since the background dilutes faster than the GW energy density (for $\wre > 1/3$ scenarios), these frequency modes are more enhanced. Thus, for $\wre > 1/3$, this bound imposes even stronger constraints on both the magnetic spectral index and reheating dynamics.

\paragraph{\underline{\bf Non-helical magnetogenesis in light of PTA:}}
The Stochastic Gravitational-Wave Background (SGWB) is a prominent prediction from numerous astrophysical and cosmological phenomena, especially within Early Universe Cosmology. Recent advancements in pulsar timing arrays (PTAs) have led to significant breakthroughs, resulting in the detection of the SGWB \cite{NANOGrav:2023gor, 2023arXiv230616224A, Reardon:2023gzh, Xu:2023wog}. Additionally, a thorough analysis of the Hellings-Downs (HD) correlation of timing residuals has provided substantial evidence for the SGWB, clarifying the observed power-law excess. Notably, the significance of the HD correlation for various PTAs falls approximately within $3\sigma$ for NANOGrav \cite{NANOGrav:2023gor}, $3\sigma$ for EPTA \cite{2023arXiv230616224A}, $2\sigma$ for PPTA, and an impressive $4.5\sigma$ \cite{Reardon:2023gzh} for CPTA \cite{Xu:2023wog} observations, respectively.

As observed, inflationary magnetogenesis models can produce sufficient GW strength to surpass vacuum production in intermediate frequency ranges. For $\ne > 0$, with appropriate reheating parameters, we can generate blue-tilted GW at intermediate frequencies. From the $\Delta N_{\text{eff}}$ bound, we found that for $\wre > 1/3$ reheating scenarios, the electrical spectral index $\ne$ must be tightly constrained to avoid excessive GW production at very high frequencies, where the GW spectrum scales as $\ogwh \propto f^{2\ne + |\nw|}$ for $\wre > 1/3$. This is illustrated in Fig.~(\ref{fig:gw_3}). Meanwhile, PTA observations indicate that the spectrum should exhibit a strong blue tilt at nano-Hz frequencies, with a GW spectrum scaling as $\ogwh\propto f^{0.91}$ for NANOGrav 15~yr data \cite{NANOGrav:2023gor}. Here we have found that for $\wre<1/3$ reheating scenarios, the GW spectrum produced from the EM field behaves as $\ogwh\propto f^{2\ne}$ for $f<\fre$ and $\ogwh(f)\propto f^{2\ne-|\nw|}$ for $f>\fre$, so to get a strongly blue tilted GW spectrum at the nano-Hz frequency range, the lower EoS $(\wre<1/3)$ is more favored compared to $\wre>1/3$ scenarios.

\begin{figure}
       \centering
      \includegraphics[width=0.45\textwidth]{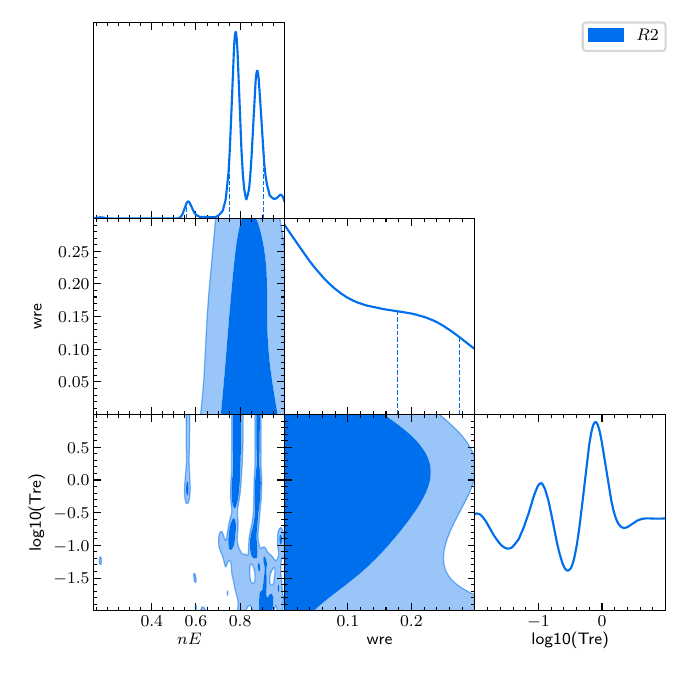}
       \includegraphics[width=0.45\textwidth]{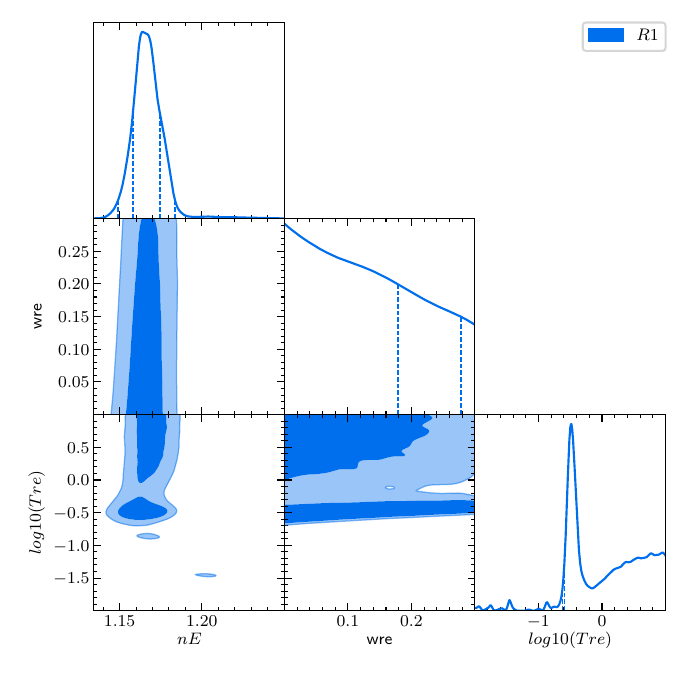}
        \caption{\textit{In the figures presented above, we display the posterior distributions of the parameters governing the Stochastic Gravitational Waves (SGWs) induced by non-helical magnetic fields, generated using the PTARcade code~\cite{2023arXiv230616377M}. The diagonal elements of the corner plot show the 1D marginalized distributions for the parameters $\ne$, $\wre$, and $\Tre$. In the left figure, we consider a prior for the electric spectral index $ \ne \in (0,1.0) $, while in the right figure, we consider an extended prior range for $\ne \in (0,1.5)$.}}
        \label{fig:mcmc_1}
    \end{figure}

For this analysis, we employed the PTARcade code \cite{2023arXiv230616377M} for Markov Chain Monte Carlo (MCMC) analysis to determine the best-fit parameter values, restricting our scenarios to $\wre < 1/3$ reheating cases and running the PTARcade code. We selected the following priors: $\wre \sim \mathcal{U}(0,1/3)$, $\ne \sim \mathcal{U}(0,1.2)$, and $\log_{10}(\Tre) \sim \mathcal{U}(-2,1)$. The predicted parameters and their $1\sigma$ tolerance range, along with the corresponding Bayesian factor, are listed in Tab.~(\ref{Tab3}). The posterior distributions of the predicted parameters are shown in Fig.~(\ref{fig:mcmc_1}) for two different runs $(R1, R2)$ corresponding to two different priors of $\ne \in (0,1.0)$ and $\ne \in(0,1.5)$ respectively. Our analysis indicates that, to match current data, this model can only produce a GW signal consistent with both amplitude and slope if $\ne \simeq 1.21^{+0.1}_{-0.12}$. Note the $\ne$ is greater than unity. 
\begin{table} 
 \begin{tabular}{|c c c c c|} 
 \hline
\large{Model} & \large{Parameter} & \large{Prior} & \large{Posterior} & \large{Bayes Factors}~($\mathcal{B}_{\rm X,Y}$)\\
\hline
\hline
\textbf{R1} & $\wre$ &  $(0,0.333)$ & $0.12^{+0.11}_{-0.08}$ &\\
               &  $\log_{10}(\Tre)$ & $(-2,1)$ & $-0.40^{+0.93}_{-1.0}$ & $1.57\pm 0.19$\\
               & $\ne$ & $(0,1.5)$ & $0.32^{+0.65}_{-0.10}$ &\\
               \hline
  \textbf{R2}   &     
  $\wre$ &  $(0,0.333)$ &  $0.13^{+0.11}_{-0.09}$ &\\
               &  $\log_{10}(\Tre)$ & (-3,1) & $-0.38^{+1.06}_{-0.17}$ & $22.77\pm 7.022$\\
               & $\ne$ & (0,1.5) & $1.17^{+0.01}_{-0.01}$ &\\
 \hline              
 \end{tabular}
 \caption{ In the table above, we list the posterior distributions of the parameters for our model, obtained from two different runs, $R1$ and $R2$. In $R1$, we consider the prior for the electric spectral index $ \ne \in (0,1.0)$, which is consistent with the $ \Delta N_{\rm eff}$ bound. In $ R2$, we extend the prior range of $\ne$ to $ (0,1.5) $, while keeping the same prior for the other two parameters in both cases. We also include the corresponding Bayes Factor for each run in the table.}
 \label{Tab3}
\end{table}
Let us remind that for $\wre < 1/3$, the  GW spectrum exhibits a blue tilt for $f > f_{\rm re}$, with $\ogwh \propto f^{2\ne - |\nw|}$. Where $\nw = 2(1-3\nw)/(1+3\nw) \leq 2$.  
Hence, for $\ne > 1.0$, the GW spectrum exhibits a blue tilt at high frequency for $f > f_{\rm re}$.  Consequently, it again violates $\Delta N_{\text{eff}}$ bound.  
Including other PTA observations, such as those from the European PTA (EPTA) (including data from the Indian PTA (InPTA) \cite{2023arXiv230616224A}, PPTA \cite{Reardon:2023gzh}, or CPTA \cite{Xu:2023wog}, may suggest a slightly lower spectral index. However, we find it difficult, or one needs extreme fine-tuning of the model parameters, to be able to fit the nano-Hz observation. 

\section{Conclusions}\label{sec5}
In this study, we have investigated the generation of primordial magnetic fields through an inflationary coupling of the form $ f^2(\phi) F_{\mu\nu}F^{\mu\nu} $, with a particular focus on the effects of non-instantaneous reheating. We explored how the electrical conductivity during the reheating phase influences the evolution of electromagnetic fields, analyzing two limiting cases: zero and infinite conductivity. Our analysis demonstrates that the reheating phase plays a critical role in determining the present-day strength of magnetic fields on large scales.

By varying the reheating parameters, namely the effective equation of state $\wre$ and the reheating temperature $\Tre$, we showed that it is possible to satisfy current observational bounds on large-scale magnetic fields within the coupling range $-2.1 < n < 0$, provided the reheating scenario involves negligible electrical conductivity. This regime is also free from strong coupling and backreaction problems. For example, under matter-like reheating ($\wre = 0$) with $\Tre \simeq 10^{-2}\,\text{GeV}$ and $n \simeq -2.1$ ($\ne \simeq -0.2$), the present-day magnetic field strength at the 1 Mpc scale can reach 
$
B_0(1\,\text{Mpc}^{-1}) \simeq 4.26 \times 10^{-25}\,\text{G}.
$
For a radiation-like evolution with the same parameters, the strength can be as high as 
$
B_0(1\,\text{Mpc}^{-1}) \simeq 3.99 \times 10^{-13}\,\text{G}.
$
In the case of stiff-like evolution ($\wre = 0.5$), the maximum field strength achievable for $\ne \simeq 0.25$ and $\Tre \simeq 10^{-2}\,\text{GeV}$ is about 
$
B_0(1\,\text{Mpc}^{-1}) \simeq 4.97 \times 10^{-15}\,\text{G}.
$
These results show that for $\wre > 1/3$, slightly blue-tilted electric spectra can produce magnetic fields strong enough to be detectable in future gravitational wave experiments.

In the second part of this work, we explored the generation of secondary gravitational waves (SGWs) from these magnetogenesis scenarios. For $\wre < 1/3$, the SGW spectrum peaks at high frequencies and could be detectable by future GW detectors. In contrast, for $\wre > 1/3$, the gravitational wave spectrum undergoes non-trivial evolution and may be relevant even at CMB scales, offering potential constraints via future B-mode polarization measurements. Most notably, higher $\wre$ values lead to multiple spectral breaks in the present-day GW spectrum, which could be observed by experiments like LISA, BBO, DECIGO, or even SKA. While radiation-like reheating allows the largest magnetic field strengths, the resulting GW signals often fall below detector sensitivity. However, higher $\wre$ scenarios are not only favorable for magnetogenesis but also offer greater promise for GW detection.

In summary, our results show that to successfully realize inflationary magnetogenesis in Ratra-type models, without encountering strong coupling or backreaction issues, one must consider non-trivial reheating scenarios with negligible electrical conductivity. Furthermore, some parameter ranges allow for the simultaneous generation of detectable gravitational wave signals. Our study, therefore, provides a crucial framework for connecting inflationary magnetogenesis with future GW observations and for constraining the physics of reheating.

 \acknowledgments
SM gratefully acknowledges financial support from the Council of Scientific and Industrial Research (CSIR), Ministry of Science and Technology, Government of India. SM also thanks Prof. L. Sriramkumar for insightful and fruitful discussions related to this work. DM wishes to acknowledge support from the Science and Engineering Research Board (SERB), Department of Science and Technology (DST), Government of India (GoI), through the Core Research Grant CRG/2020/003664.  We want to thank our Gravity and High Energy Physics groups at IIT Guwahati for illuminating discussions.
\appendix
\section{Calculation of the coefficient  $\mathcal{A}_1,~\mathcal{A}_2~\&~\mathcal{A}_3$:}\label{Appendix}

The tensor power spectrum sourced by electromagnetic fields at the end of reheating is given by
\begin{align}\label{eq:ptres_a1}
    \Ptsre=\frac{2\HI^4}{\Mp^4}\left(\frac{k}{\ke}\right)^{2(\delta-2)}\mCbe^2(\nbe)\left(\frac{k}{\kf}\right)^{2\nbe}\Cmre^2(\xre,\xe)\fnbe(k),
\end{align}
where the time-dependent integral part $\Cmre(\xre, \xe)$ is defined as \cite{Maiti:2024nhv}
\begin{align}
    \Cmre(\xre,\xe)=\int_{\xe}^{\xre}dx_1\,x_1^{-\delta}\mGk(\xre,x_1),\label{eq:cmre_a}
\end{align}
with $\mGk(x, x_1)$ being the Green’s function associated with tensor perturbations. During the reheating era, this Green’s function takes the form~\cite{Maiti:2024nhv}
\begin{align}
    \mGk(x,x_1) 
= \theta(x-x_1)\frac{\pi x^{l} x_1^{1-l}}{2k\mathrm{sin}(l\pi)}
\l[J_l(x)J_{-l}(x_1)-J_{-l}(x)J_l(x_1)\r]\label{eq:gk_a}
\end{align} 
Substituting Eq.\eqref{eq:gk_a} into Eq.\eqref{eq:cmre_a}, we obtain
\begin{align}\label{D2}
\Cmre(x,x_1)
&
=\f{\pi x^l}{2\mathrm{sin}(l\pi)}
\int d x_1x_1^{1-l-\delta}\l[J_l(x)J_{-l}(x_1)-J_{-l}(x)J_l(x_1)\r]\nn\\
&=2^{-2-l} x^l x_1^{2-2l-\delta}
\biggl\{\f{\Gamma(-l)\Gamma[1-(\delta/2)]}{\Gamma[2-(\delta/2)]}
x_1^{2l}J_{-l}(x)\,
{}_1F_2[1-(\delta/2); 1+l,2-(\delta/2);-(x_1^2/4)]\nn\\
& +4^l \f{\Gamma(l)\Gamma[1-l-(\delta/2)]}{\Gamma[2-l-(\delta/2)]}J_l(x)
{}_1F_2[1-l-(\delta/2);1-l,2-l-(\delta/2);-(x_1^2/4)]\biggr\},
\end{align}
where ${}_1F_2(a,b,c,z)$ denotes the hypergeometric function.

\subsubsection{Spectral shape of $\Omega_{_{\mathrm{GW}}}$
for $k<\kre$ and $k\gg\kre$}

For wave numbers satisfying $ k \ll \kre $, we consider the regime where $\xe \ll \xre \ll 1$.
In this limit, the integral $\Cmre(\xre, \xe)$ simplifies significantly and can be approximated as
\begin{align}\label{eq:cm1}
\lim_{k\ll\kre} \Cmre(\xre,\xe)
\simeq\f{1}{(1-\delta)^2} 
\l\{\f{2}{1+2\delta}
-\f{2}{2-\delta} 
\l[1-\l(\f{\kre}{\ke}\r)^{2-\delta}\r]\r\}
\l(\f{k}{\kre}\r)^{2-\delta}.
\end{align} 
Similarly, when $k\gg\kre$, for $\wre>1/3$, the quantity $\Cmre(\xre,\xe)$ 
reduces to
\begin{align}\label{eq:cm2}
\lim_{k\gg\kre}  
\Cmre(\xre,\xe)
&\simeq \sqrt{\f{2}{\pi}}\f{2^{(1-\delta)/2}}{(1-\delta)}
\Gamma[(3-\delta)/2] \Gamma[(\delta-1)/2]\xre^{-\delta/2}
\biggl\{\f{\Gamma[(1-\delta)/2]}{\Gamma(\delta/2)}
\cos\l[\xre-(2-\delta)\pi/4\r]\nn\\ 
& - \frac{2}{2-\delta} 
\cos\l[\xre-\delta\pi/4\r]\biggr\},
\end{align}
whereas, for $\wre<1/3$, the quantity simplifies to
\begin{align}\label{eq:cm3}
\lim_{k\gg\kre} \Cmre(\xe,\xre)
&\simeq  \sqrt{\f{2}{\pi}}\f{2^{(\delta-3)/2}}{(1-\delta)}
\f{\Gamma[(\delta-1)/2]\Gamma[1-(\delta/2)]}{\Gamma[2-(\delta/2)]}
\xre^{-\delta/2}\xe^{2-\delta}  \cos\l[-\xre+\delta\pi/4\r]\nn\\
&\simeq \sqrt{\frac{2}{\pi}}
\frac{2^{(\delta-3)/2}\Gamma[\frac{\delta-1}{2}]}{(1-\delta)(1-\delta/2)}
\xre^{-\delta/2}\xe^{2-\delta}\cos[\xre-(2-\delta)\pi/4].
\end{align}

  At the end of reheating, the power spectrum of secondary GWs generated due to the magnetic fields [defined in Eq.~\eqref{eq:ptres}] is given by
\begin{align}\label{eq:ptres_a}
    \Ptsre=\frac{2\HI^4}{\Mp^4}\left(\frac{k}{\ke}\right)^{2(\delta-2)}\mCbe^2(\nbe)\left(\frac{k}{\kf}\right)^{2\nbe}\Cmre^2(\xre,\xe)\fnbe(k),
\end{align}
On substitutingEq.~\eqref{eq:cm1} in this expression, for $k\ll \kre$, we obtain that
\begin{align}
\lim_{k<<\kre}\Ptsre(k,\ere)\simeq \mA_1\,\l(\frac{\HI}{\Mpl}\r)^4\l(\f{\kre}{\ke}\r)^{2(\delta-2)} \left(\frac{k}{\kf}\right)^{2\nbe}\,\mCbe^2(\nbe)\fnbe(k)
\end{align}
Similarly, on utilizing Eqs.~\eqref{eq:cm2} and \eqref{eq:cm3}, 
for $k > \kre$, we obtain the following expressions
\begin{subequations}
\begin{align}
\lim_{k>>\kre}\Ptsre(k,\ere) 
&\simeq \mathcal{A}_2 
\l(\f{\HI}{\Mpl}\r)^4 \l(\f{\kre}{\ke}\r)^{2(\delta-2)}
\l(\frac{k}{\kre}\r)^{-2-|\nw|}\l(\frac{k}{\ke}\r)^{2\nb} \mCbe^2(\nbe)\,\fnbe(k);\,\, \text{for}\,\, \wre>1/3\\
\lim_{k>>\kre}\Ptsre(k,\ere) 
&\simeq \mathcal{A}_3
\l(\f{\HI}{\Mpl}\r)^4 \l(\f{\kre}{\ke}\r)^{2(\delta-2)}
\l(\frac{k}{\kre}\r)^{-2-|\nw|}\l(\frac{k}{\ke}\r)^{2\nb}\mCbe^2(\nbe)\, \fnbe(k);\,\, \text{for}\,\, \wre<1/3,
\end{align}
\end{subequations}
where we defined the quantities $\mathcal{A}_1$, $\mathcal{A}_2$ and $\mathcal{A}_3$ 
that appear in the above expressions are given by
\begin{subequations}
\begin{align}
\mathcal{A}_1 
&=\f{2}{(1-\delta)^4} 
\l\{\f{2}{1+2\delta}
-\f{2}{2-\delta} 
\l[1-\l(\f{\kre}{\ke}\r)^{2-\delta}\r]\r\}^2,\\
\mathcal{A}_2 
&=\f{2^{(3-\delta)}\Gamma^2[(3-\delta)/2]
\Gamma^2[(\delta-1)/2]}{\pi(1-\delta)^2}
\biggl\{\f{\Gamma[(1-\delta)/2]}{\Gamma(\delta/2)}
\cos\l[(k/\kre)-(2-\delta)\pi/4\r] 
-\f{2}{2-\delta} 
\cos\l[(k/\kre)-\delta\pi/4\r]\biggr\}^2,\\
\mathcal{A}_3 
&=\f{2^{\delta-1}\Gamma^2[(\delta-1)/2]}{\pi(1-\delta)^2(1-\delta/2)^2}
\l(\f{\kre}{\ke}\r)^{2(2-\delta)}\cos^2[(k/\kre)-\delta{\pi}/{4}].
\end{align}
\end{subequations}

\bibliographystyle{apsrev4-1}
\bibliography{references}

\end{document}